\pdfoutput=1

\documentclass[twocolumn]{aastex63}
\usepackage{epstopdf}
\usepackage{lineno}
\usepackage{xspace}
\usepackage{graphicx}
\usepackage{amssymb}
\usepackage{amsmath}
\usepackage{natbib}
\usepackage{float}
\usepackage{hyperref}

\shorttitle{Soft X-ray Structures in Cen\,A Galaxy}
\shortauthors{Kr\'ol et al.}

\begin{document}

\title{An Analysis of Soft X-ray Structures\\ at Kiloparsec Distances from the Active Nucleus of Centaurus\,A Galaxy}

\correspondingauthor{Dominika {\L}. Kr{\'o}l}
\email{dominika.lucja.krol@gmail.com}

\author{Dominika {\L}. Kr{\'o}l}
\affiliation{Astronomical Observatory, Jagiellonian University, ul. Orla 171, 30-244 Krak\'ow, Poland}

\author{Volodymyr Marchenko}
\affiliation{Astronomical Observatory, Jagiellonian University, ul. Orla 171, 30-244 Krak\'ow, Poland}

\author{Micha{\l} Ostrowski}
\affiliation{Astronomical Observatory, Jagiellonian University, ul. Orla 171, 30-244 Krak\'ow, Poland}

\author{{\L}ukasz Stawarz}
\affiliation{Astronomical Observatory, Jagiellonian University, ul. Orla 171, 30-244 Krak\'ow, Poland}

\begin{abstract}
Here we re-analyze the archival {\it Chandra} data for the central parts of the Centaurus\,A radio galaxy, aiming for a systematic investigation of the X-ray emission associated with the inner radio lobes in the system, and their immediate surroundings. In particular, we focus on four distinct features characterized by the soft excess with respect to the adjacent fields. Those include the two regions located at kpc distances from the nucleus to the West and East, the extended bow-shock structure to the South, and a fragment of a thin arc North from the center. The selected North, West, and South features coincide with the edges of the radio lobes, while the East structure is seemingly displaced from the radio-emitting plasma. Our X-ray spectral analysis reveals (i) a power-law emission component with photon index $\Gamma \sim 2$ in the North, East, and South regions, and (ii) a dense (number density $\sim 0.3$\,cm$^{-3}$) and relatively cold (temperature $\sim 0.2$\,keV) gas in the East and West regions. The power-law emission is consistent with the synchrotron continuum generated at the edges of the radio structure, and implies that the efficiency of the electron acceleration at the terminal bow-shock does not vary dramatically over the inner lobes' extension. The presence of gaseous condensations, on the other hand, could possibly be understood in terms of a massive outflow from the central regions of the galaxy.
\end{abstract}


\section{Introduction} \label{S:intro}

The famous radio galaxy Centaurus\,A, hosted by NGC\,5128, is located at the distance of $3.85 \pm 0.35$\,Mpc \citep{Rejkuba04,Ferrarese07}. NGC\,5128 is an elliptical with a prominent dust lane and other morphological features indicating multiple merger events which occurred  about 200--700 million years ago \citep[see the reviews by][and references therein]{Israel98,Morganti10}. It harbors a central black hole with the estimated mass of $(0.5-1.1) \times 10^8 M_{\odot}$ \citep{Marconi06,Neumayer07,Cappellari09}. Cen\,A is a strong  source of a multi-frequency emission detected on various scales. The large scale structure of the  system has been resolved in the radio domain, from very low frequencies up to mm wavelengths, to have an angular size of $8^{\circ} \times 4^{\circ}$, equivalent to the projected linear dimensions of $500$\,kpc\,$\times 250$\,kpc \citep[][]{Hardcastle09,Feain11,McKinley13}. These giant lobes have been also resolved in high-energy $\gamma$-rays by the Large Area Telescope (LAT) onboard the {\it Fermi} satellite \citep{Abdo10a,Sun16}, and selectively mapped in X-rays with {\it Suzaku} \citep{Stawarz13}. 

 On smaller scales, in particular at the distance of several--to--tens of kpc North from the Cen\,A nucleus, a diffuse and low-surface brightness radio structure called the``northern middle lobe'' is seen \citep{Morganti99}. Around and within this structure, a complex net of optical filaments of ionized gas, clouds of atomic gas with anomalous velocities, young stars, and large-scale X-ray filaments  composed of discrete knots, have been observed, all suggestive of a complex interaction between the evolving large-scale radio jet with the interstellar medium \citep{Morganti99,Oosterloo05,Kraft09,Crockett12,Neff15,Salome16}.

\begin{figure*}[!t]
\centering
\includegraphics[width=0.49\textwidth]{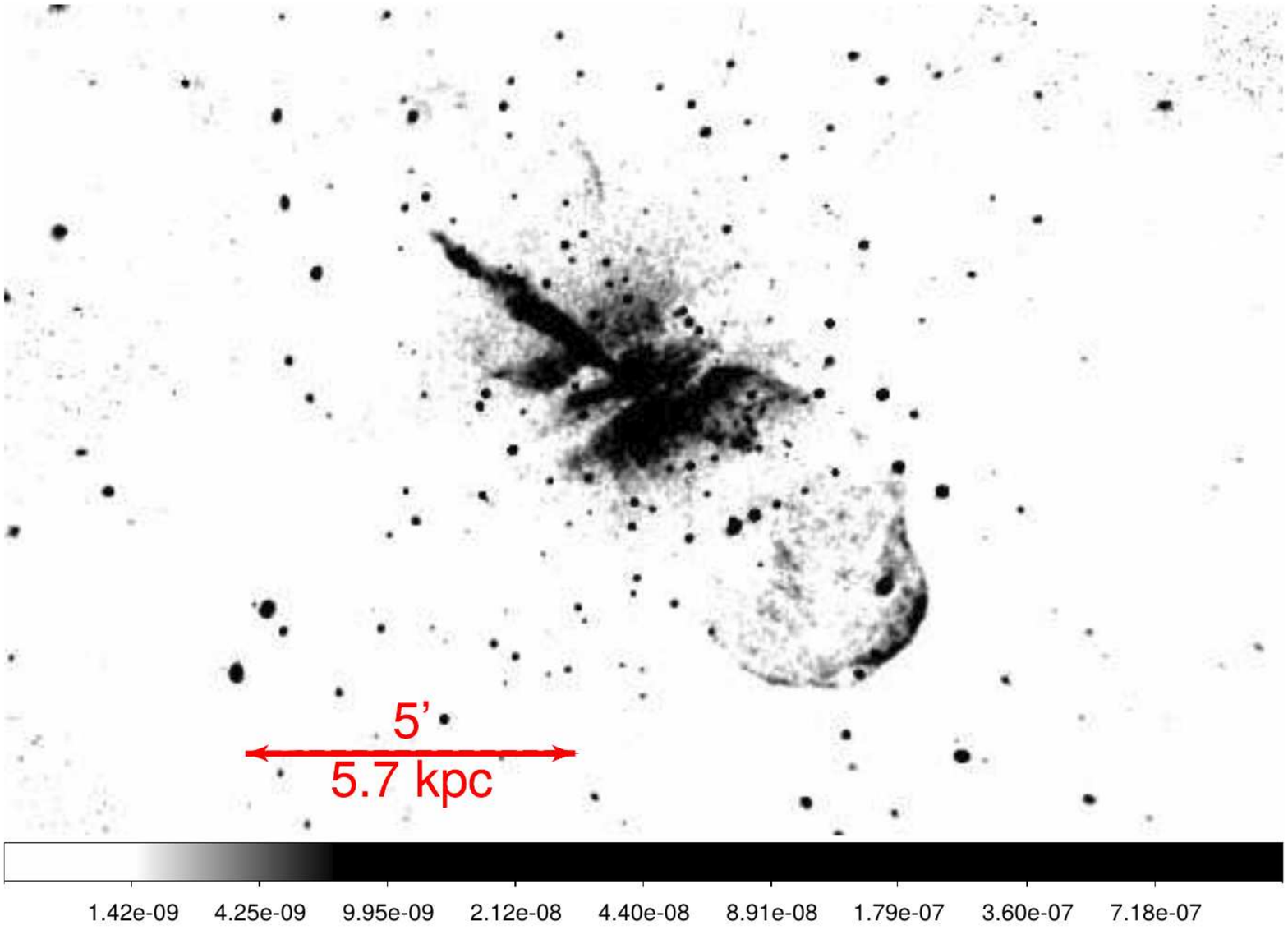}
\includegraphics[width=0.49\textwidth]{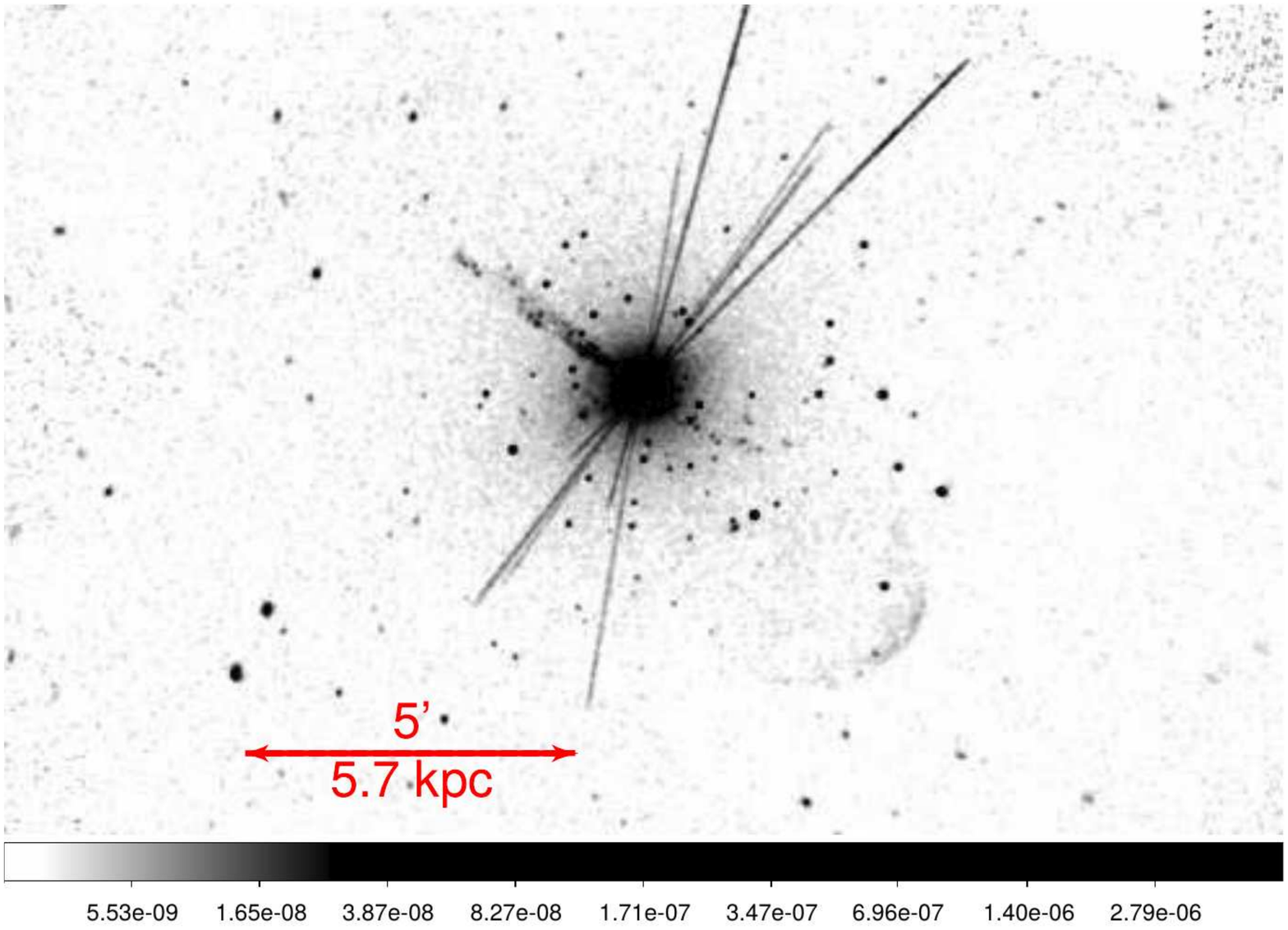}
\caption{The {\it Chandra} ACIS intensity map of the inner parts of Cen\,A, within the photon energy range $0.4 - 2.5$\,keV (left panel) and $2.7-8.0$\,keV (right panel), for the merged selected ObsIDs (316, 962, 2978, 3965, 8489 and 8490), smoothed with the $3\sigma$ Gaussian.}
\label{fig:xray_map}
\end{figure*}
	
Finally, on yet a smaller scale of a few/several arcmin, the inner structure of Cen\,A radio galaxy consists of several components clearly visible in radio and X-rays when imaged with the arcsec resolution, including the bright nucleus, the jet extending to the North-East up to four kpc from the core, and the counter-lobe pronounced to the South \citep[e.g.,][]{Hardcastle07,Croston09}. The X-ray emission spectrum of the nucleus in the energy range $3-10$\,keV is well fitted by a heavily absorbed power-law model plus a neutral and narrow fluorescence iron line \citep{Evans04}; the jet X-ray emission continuum, contributed by multiple bright knots and a diffuse component, is best described as an unabsorbed steep power-law  \citep[see][]{Kataoka06,Snios19}.  Another X-ray feature within the inner parts of the Cen\,A system, is a ring-like structure  extending to several kpc in the direction perpendicular to the jet, as reported by \citet{Karovska02}.	

The Cen\,A core has been detected in soft and high-energy $\gamma$-rays by all the instruments onboard the Compton Gamma Ray Observatory \citep[][and references therein]{Steinle98}, as well as by the {\it Fermi}-LAT \citep{Abdo10b}. The radio galaxy has also been detected in the very high energy $\gamma$-ray range ($>100$\,GeV) by the H.E.S.S. observatory \citep{Aharonian09}. The most recent analysis of the broad-band $\gamma$-ray continuum of the source reveals a spectral hardening above the photon energies of a few GeV \citep{Sahakyan13,Abdalla18}; the extension of the H.E.S.S. source reported recently by \citet{Sanchez18}, seems to point out the kpc-scale jet as the most likely origin of this observed ``excess'' $\gamma$-ray emission \citep[see, e.g.,][]{Tanada19}.

In this paper we re-analyze the archival {\it Chandra} data for the central parts of the Cen\,A radio galaxy, focusing on the spectral analysis for the diffuse features  associated with the inner radio lobes, and characterized by the excess soft X-ray emission with respect to the adjacent fields. The data acquisition, analysis, and modeling is described in \S\,\ref{S:data}; the interpretation and the discussion of the obtained results are given in \S\,\ref{S:discussion} and summarized in \S\,\ref{S:conclusions}.

\section{{\it Chandra} Data}
\label{S:data}
	
We have reviewed all the available {\it Chandra} Advanced CCD Imaging Spectrometer (ACIS) data for the innermost region of the Cen\,A system, and selected the exposures for which the ACIS readout streaks are restricted (as much as possible) to the plane perpendicular to the jet axis, in order to avoid any overlaps with the lobes and their immediate surroundings. The ObsIDs of thus selected observations are: 316, 962, 2978, 3965, 8489 and 8490. For these, the analysis was carried out with the software package {\tt CIAO\,4.10} \citep{Fruscione06} and the calibration database {\tt CALDB\,4.7.9}. Before the analysis, the data was reprocessed using the {\tt chandra\_repro} script recommended in the {\tt CIAO} analysis threads. Next the data was merged and binned with a factor of 1.0, which corresponds to the original {\it Chandra} pixel size of $0.492\arcsec$. The images of the selected observations, merged and smoothed with $3 \sigma$ Gaussian, are shown in the Figure\,\ref{fig:xray_map} for the energy ranges $0.4 \, - \, 2.5$\,keV and $2.7 \, - \, 8.0$\,keV.
	
\begin{figure}[!t]
\centering
\hbox{\hspace{-4.0em} \includegraphics[width=1.25\columnwidth]{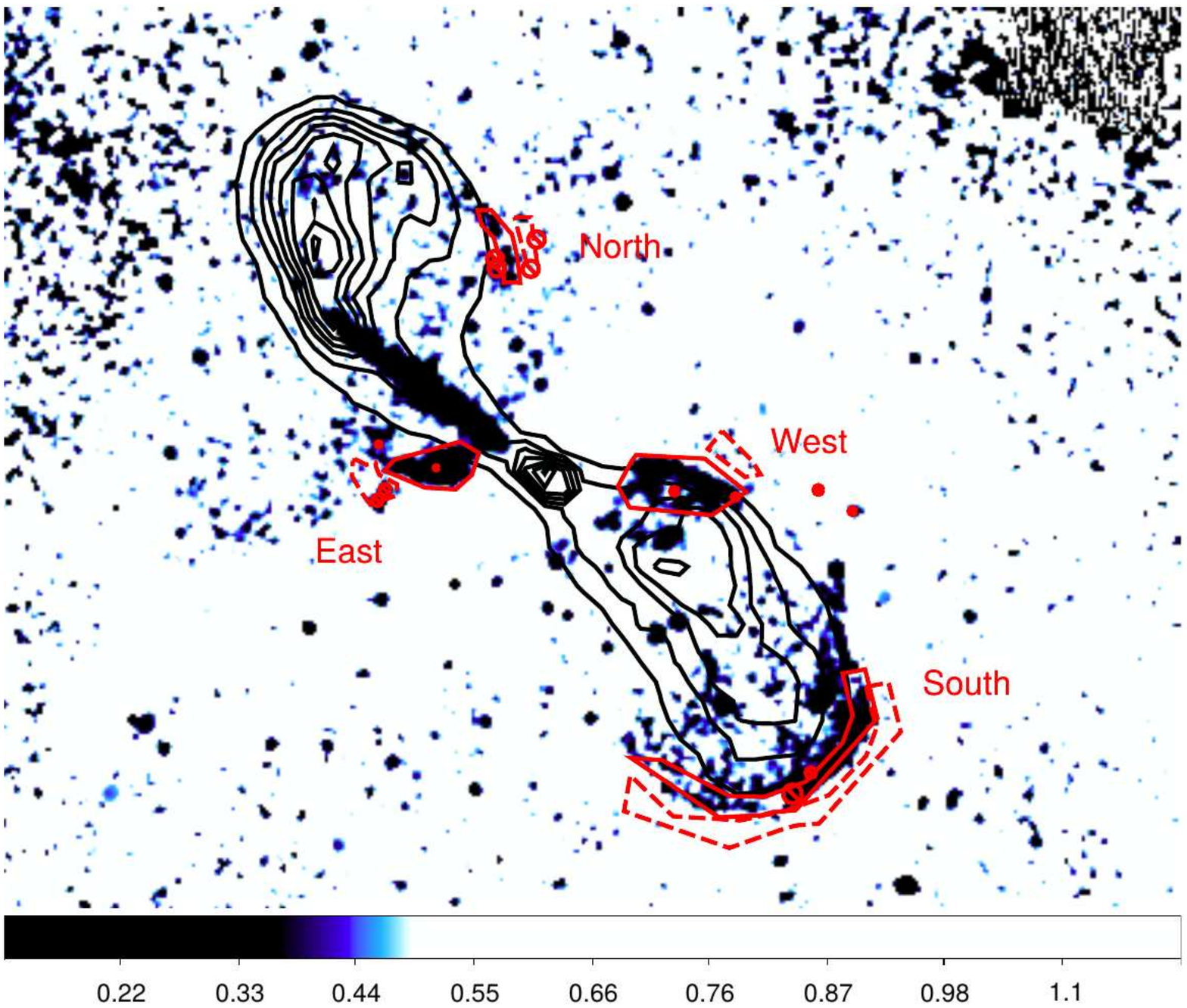}}
\caption{The smoothed (Gaussian of $\sigma = 5$\,px) hardness ratio map for the inner parts of the Cen\,A radio galaxy, with the radio continuum intensity contours superimposed (black), starting from the 0.2\,Jy/beam level, and increasing by a factor of $\sqrt{2}$.  The regions selected for the spectral analysis --- East, West, South and North --- are denoted with solid red contours, and the corresponding background regions by dashed red contours. Also marked are the point sources (mostly foreground XRBs) removed from the spectral extraction regions.}
\label{fig:reg}
\end{figure}

\subsection{X-ray hardness analysis}
\label{S:HRM}

The {\it Chandra} map of Cen\,A radio galaxy in the energy range $0.4 \, - \, 2.5$\,keV (Figure\,\ref{fig:xray_map} left), reveals a pronounced soft X-ray emission from the central kpc region, characterized roughly by an ``hour glass'' appearance, and coinciding roughly with the innermost segments of the radio lobes; this structure is not prominent in the higher energy range $2.7 \, - \, 8.0\, $\,keV (right panel of the figure). The kpc-scale jet to the North-East, and the bow-shock structure to the South marking the outer edge of the counter-lobe, are also manifesting clearly in the soft X-ray range, although both structures can also be noted on the hard X-ray {\it Chandra} map. In addition to those, in the soft image, a fragment of a thin but distinct arc located to the North from the nucleus is clearly visible as well; its position and orientation are both consistent with the extension of the edge of the main radio lobe.

 Figure\,\ref{fig:reg} presents the hardness ratio map of the analyzed system, based on the exposure-corrected images, and smoothed with the Gaussian of $\sigma = 5$\,px. The hardness ratio here is defined as the ratio of the $2.7 - 8.0$\,keV and $0.4 - 2.5$\,keV count rates, with the energy gap between the two photon energy ranges chosen to account for the energy resolution of the {\it Chandra}'s ACIS (which is of the order of 150\,eV). For comparison, in the figure we also display the contours of the lobes' continuum radio emission at 21\,cm, taken from the NRAO\footnote{The National Radio Astronomy Observatory is a facility of the National Science Foundation operated under cooperative agreement by Associated Universities, Inc.} Very Large Array (VLA) archive \citep{Condon96}. As shown in the figure, while the jet, the Southern bow-shocks, and the Northern arc, are all still prominent on the hardness map, the soft diffuse X-ray hour-glass structure ``collapses'' to two distinct but symmetric features with relatively sharp and well defined boundaries, located into the East and into the West of the core. Moreover while the aforementioned North, West, and South regions do overlap well with the lobes' edges on the radio map, the East feature appears seemingly displaced from the radio-emitting plasma of the main lobe.

 We note that the two peculiar structures East and West are inclined at $\sim 35$\,deg with respect to the jet axis, and are approximately cone-like shaped; the core-vertex angular separation for both reads as $\sim 2.4$\,arcmin each, meaning the projected distances of about 2.7\,kpc for the conversion scale 19\,pc/arcsec adopted hereafter.
	
 The exact regions selected for the following spectral analysis are shown in Figure\,\ref{fig:reg}. The corresponding background regions were chosen to avoid any overlaps with the radio lobes on one hand, and the dust lane on the other hand (for the East and West features); we emphasize, however, that the choice of the background region in such a crowded field deep within the host galaxy and close to the very bright nucleus, is in general rather difficult, and will always be, to some extent, arbitrary. We have therefore repeated the fitting with the enlarged and modified background regions, obtaining consistent results. We have also attempted an analysis of the spectra of the other segments of the lobes' edges/surroundings, including mirror reflections of the West and East regions with respect to the jet axis, but due to a combination of a complicated field and low photon statistic, we failed to obtain proper fits for such.

 We finally note that, the X-ray emission of the regions overlapping with the South and West features studied in this paper, was examined before by \citet{Croston09}; the East and North features, however, were never subjected to any spectral modeling in the past. Also, here we do not discuss the kpc-scale jet emission, referring instead to the previous detailed analysis by \citet{Kataoka06} for the jet diffuse X-ray emission component, and \citet{Snios19} for the most recent update on the compact jet knots. 
	
\begin{figure*}[!t]
	\centering
	\includegraphics[width=0.49\textwidth]{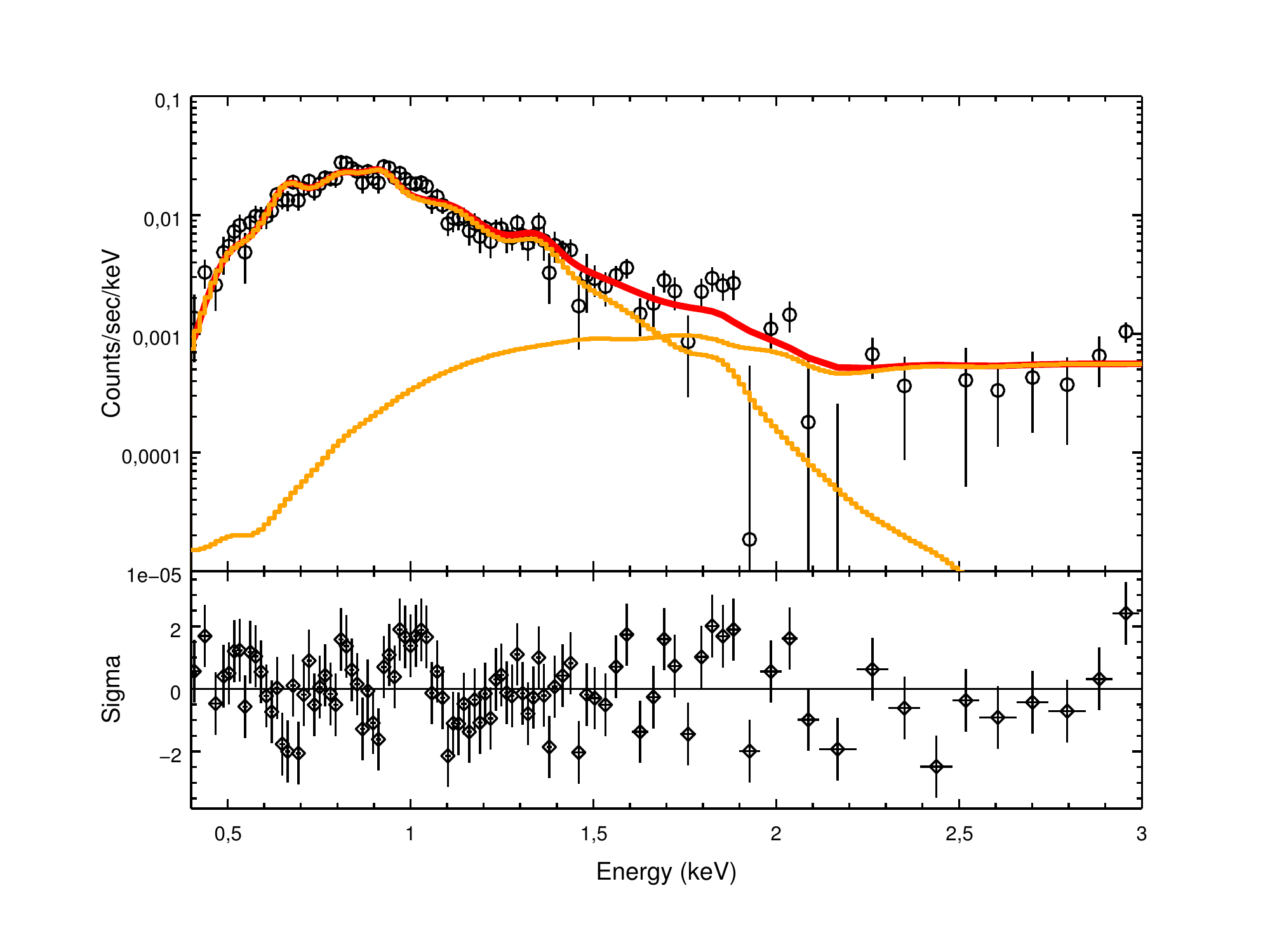}
	\includegraphics[width=0.49\textwidth]{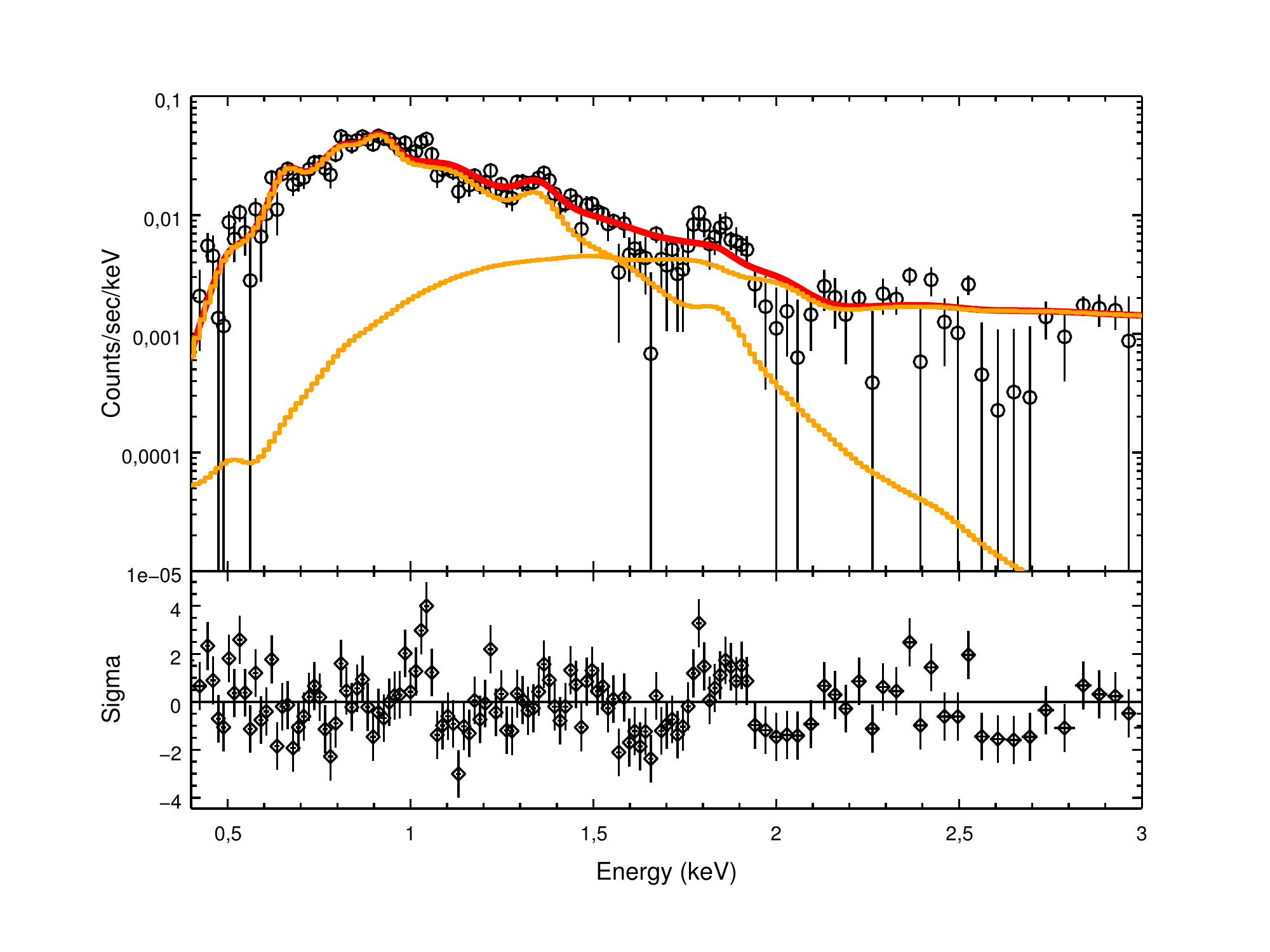}
	\includegraphics[width=0.49\textwidth]{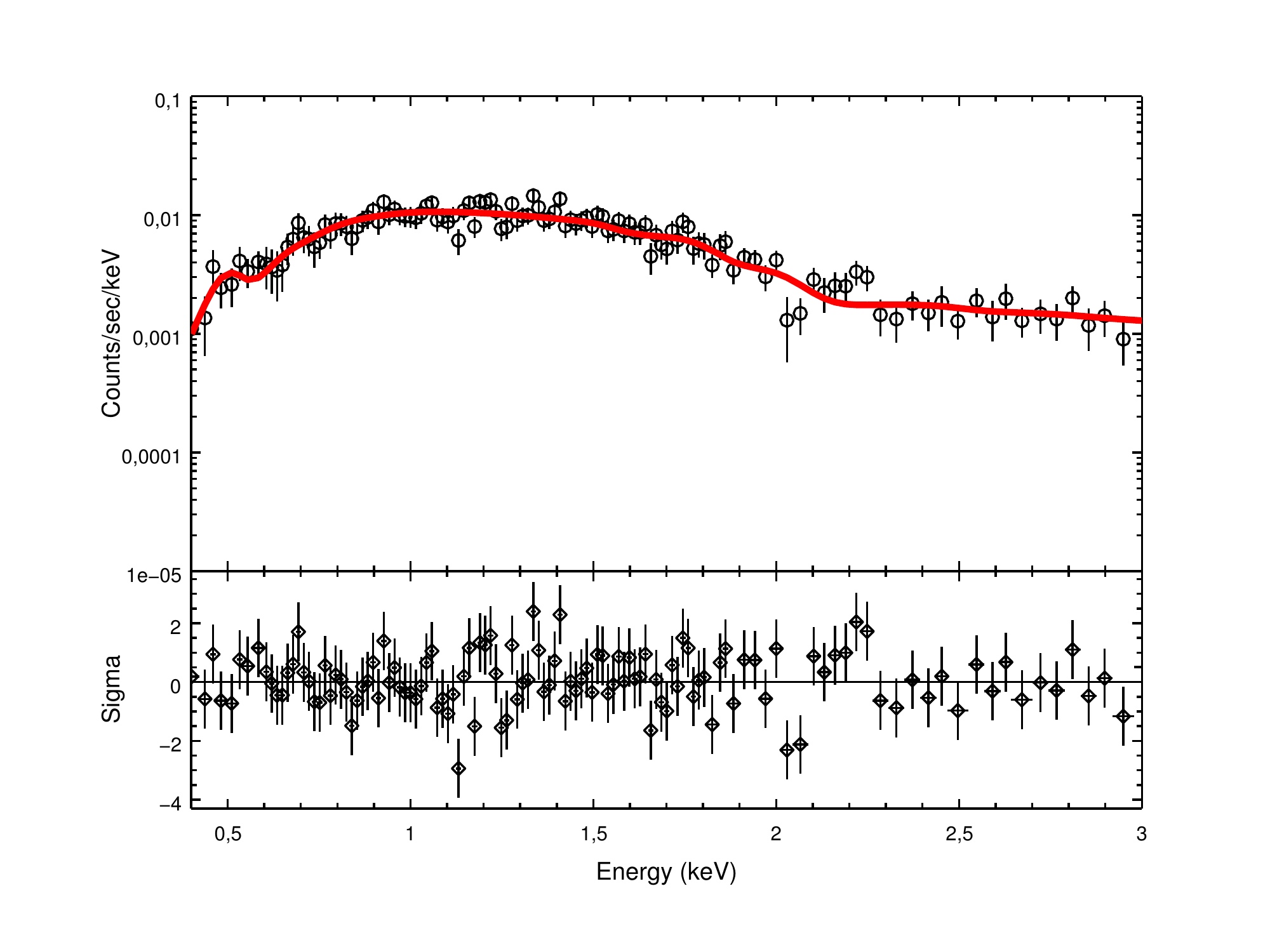}
	\includegraphics[width=0.49\textwidth]{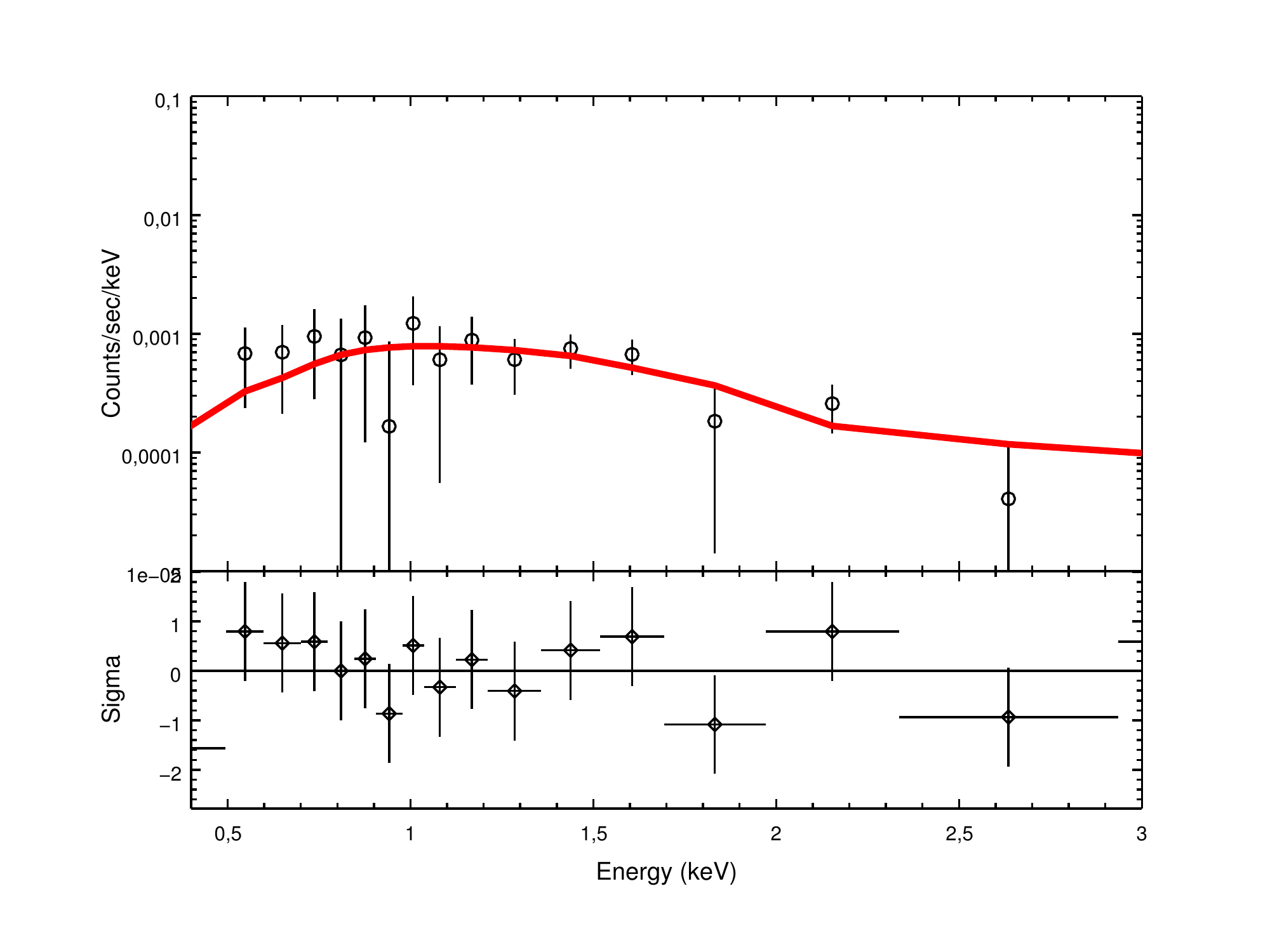}
	\caption{ {\it Chandra} spectra along with the best-fit models (and residuals) for the four regions selected for the analysis: East (upper left), West (upper right), South (lower left), and North (lower right). The models displayed consist of a mixture of absorbed thermal and non-thermal components, {\tt xsphabs*(xsapec+xspowerlaw)}, for the West and East regions, and absorbed power-law, {\tt xsphabs*xspowerlaw}, for the North and West regions.}
	\label{fig:fit_dw}
\end{figure*}

\subsection{Spectral modelling}

For the spectral analysis of the selected regions, first we extracted the corresponding spectra from each ObsID considered in our study, using the {\tt specextract} tool from the {\tt CIAO} software package. Next we combined the spectra for each region using the {\tt combine\_spectra} script. Because of the extended wings of the point spread function (PSF) of the extremely bright nucleus, which in addition is subjected to a severe pile-up in the instrument, we restricted the analysis to the photon energy range up to 3\,keV \citep[see in this context also][]{Croston09}. We note, that both the East and West regions are located at similar distances from the core, and so are their background regions, and hence any photon leakage from the central PSF should affect the spectral analysis of  both regions in a similar/comparable manner;  the North and especially the South regions, on the other hand, are expected to be subjected to a much lesser extent by the pollution from the core emission.

	\begin{table*}[!t]
		{\scriptsize
			\begin{center}
				\caption{Spectral fitting results}
				\label{tab:fit}
				\begin{tabular}{cllll}
					\hline\hline
					Region/Model &  Parameter & Value &  $1\sigma$ errors & Units \\ 
					\hline\hline\\
					{\bf East}  & kT  &  0.23 & 0.01& keV \\ 
					{\tt xsphabs*(xsapec+xspowerlaw)} & norm  & $0.4$ & $0.1$ & $10^{-2}\times $\,{\tt apec} \\ 
					& Abundanc  & 0.09 & 0.03& --- \\         
					& $\Gamma$  & 0.24 & 0.66 & --- \\ 
					& ampl  & $2.5$ & $1.5$ & $10^{-6} \times$\,ph/keV/cm$^2$ at 1keV \\  
					& $N_{\rm H}$  & 0.40 & 0.04& $10^{22}$\,cm$^{-2}$\\ 
					& Final fit statistic & 125.33 & &   \\
					& Degrees of freedom  & 92 & & \\
					\\
					\hline
					\hline\\
					{\bf West} & kT &  0.19& 0.01& keV \\ 
					{\tt xsphabs*(xsapec+xspowerlaw)} & norm & $1.9$ & $0.7$ & $10^{-2}\times $\,{\tt apec} \\ 
					& Abundanc  & 0.20 & 0.09& --- \\ 
					& $\Gamma$  & 1.75&  0.28& --- \\ 
					& ampl  & $32.1$ & $7.1$ & $10^{-6} \times$\,ph/keV/cm$^2$ \text{at } 1 keV \\ 
					& $N_{\rm H}$  & 0.65 & 0.03& $10^{22}$\,cm$^{-2}$\\ 
					& Final fit statistic & 208.34 & &  \\
					& Degrees of freedom  & 128 &  & \\
					\\
					\hline
					\hline\\
					{\bf South} & kT &  0.52& 0.03& keV \\ 
					{\tt xsphabs*(xsapec+xspowerlaw)} & norm & $0.06$ & $0.01$ & $10^{-2}\times $\,{\tt apec} \\ 
					& Abundanc  & $0$ & (unconstr.) & --- \\ 
					& $\Gamma$  & 1.00&  0.20& --- \\ 
					& ampl  & $15.2$ & $3.3$ & $10^{-6} \times$\,ph/keV/cm$^2$ at 1keV \\ 
					& $N_{\rm H}$  & 0.16 & 0.03& $10^{22}$\,cm$^{-2}$\\ 
					& Final fit statistic &179.68& &   \\
					& Degrees of freedom  & 118 & & \\
					\hline
					{\tt xsphabs*xspowerlaw} & $\Gamma$  & 2.14 & 0.13 & --- \\ 
					& ampl  & $46.4$ & $4.2$ &  $10^{-6} \times$\,ph/keV/cm$^2$ at 1keV \\  
					& $N_{\rm H}$  & $0.026$ & $^{+0.030}_{- \,\, \text{---}}$ & $10^{22}$\,cm$^{-2}$\\ 
					& Final fit statistic &104.49& &   \\
					& Degrees of freedom  & 121 & & \\			
										\\
					\hline
					\hline\\
					{\bf North} & kT &  0.16& 0.05& keV \\ 
					{\tt xsphabs*(xsapec+xspowerlaw)} & norm & $8 \times 10^{-5}$ &  (unconstr.)  & $10^{-2}\times $\,{\tt apec} \\ 
					& Abundanc  & $0.24$ & (unconstr.) & --- \\ 
					& $\Gamma$  & 2.52 & 0.71 & --- \\ 
					& ampl  & $4.5$ & $2.0$ &  $10^{-6} \times$\,ph/keV/cm$^2$ at 1keV \\  
					& $N_{\rm H}$  & 0.21 &  $^{+0.26}_{- \,\, \text{---}}$ & $10^{22}$\,cm$^{-2}$\\ 
					& Final fit statistic &43.69&  &  \\
					& Degrees of freedom  & 59 & &\\
					\hline
					{\tt xspowerlaw} & $\Gamma$  & 2.02& 0.43 & --- \\ 
					 & ampl  & $3.0$ & $0.65$ &  $10^{-6} \times$\,ph/keV/cm$^2$ at 1keV \\   
					& Final fit statistic &8.82& &  \\
					& Degrees of freedom  & 15 &  & \\
					\\
					\hline\hline
		\end{tabular}
			\end{center}
		}
	\end{table*}

 After the background extraction, the fitting was performed for the grouped data (with the minimum signal-to-noise ratio $=5$ for the West, East and South region, and minimum signal-to-noice ratio $=7$ for the North region because of a much lower photon statistics) using the {\tt Sherpa} fitting application \citep{Freeman01}. The initial values of the model free parameters were chosen based on the preliminary fitting using the Monte-Carlo method in {\tt Sherpa}. 

The fitted model for all the regions included a thermal component ({\tt xsapec}) plus a power-law component ({\tt xspowerlaw}), moderated by the Galactic absorption and the internal absorption ({\tt xsphabs}). All the model parameters were set free, except of the Galactic equivalent hydrogen column density in the direction of the source, which was frozen at $N_{\rm H,\,Gal} =  7.31\times10^{20}$\,cm$^{-2}$ following \citet{Kalberla05}. For the North and South regions, we applied a simple absorbed power-law model, with no contribution from a thermal component; in fact for these regions the two-component model (power-law plus apec) did not provide any substantial improvement over the one-component model (consisting of a single power-law emission). The spectra of the East and West regions, on the other hand, could not be fitted at all with a one-component model, consisting of either an absorbed single power-law emission, or an absorbed single-temperature plasma, and hence below we do not discuss those attempts. 

\begin{figure*}[!t]
	\centering
 \includegraphics[trim=0 0 0 12cm, width=\columnwidth]{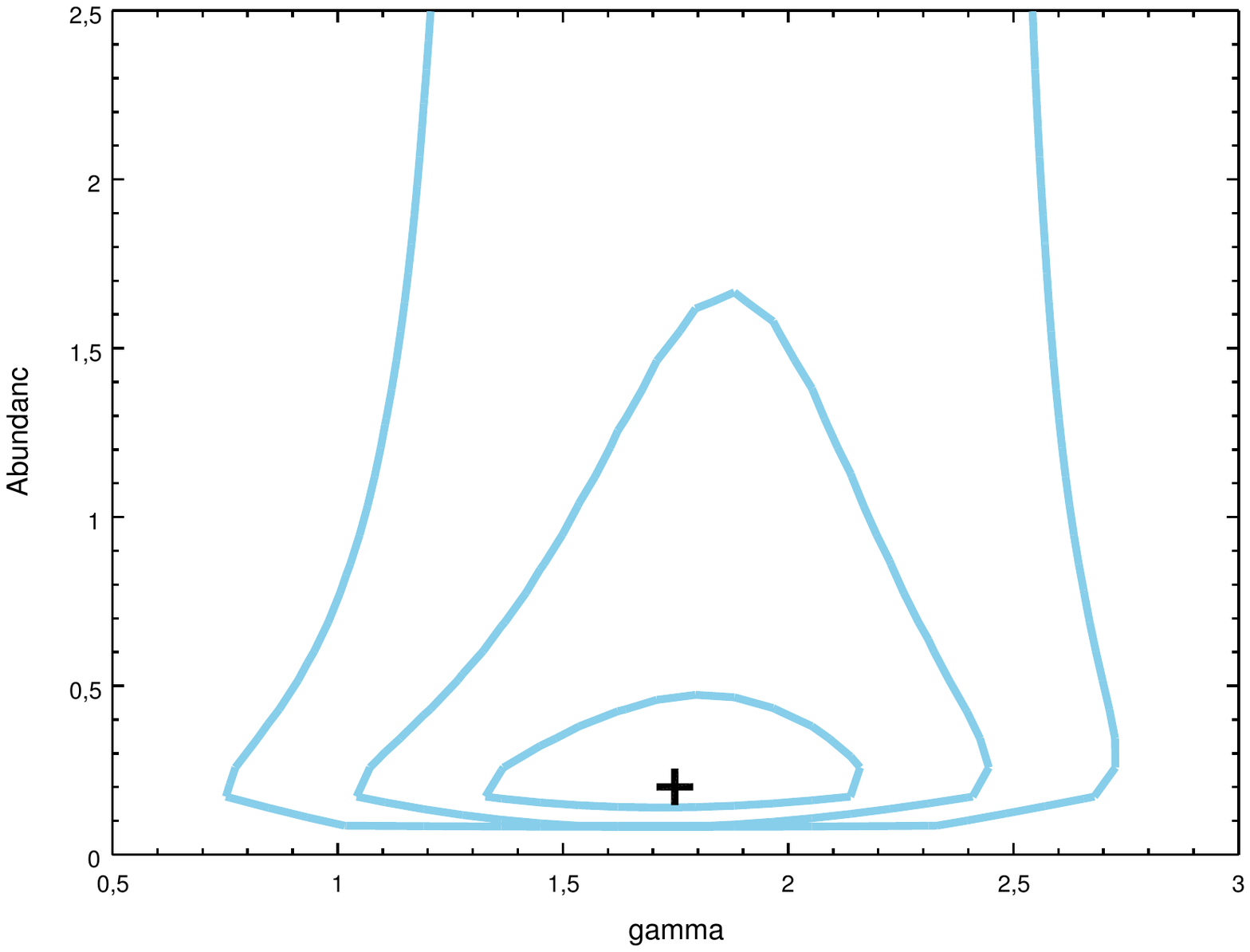}
\includegraphics[trim=0 0 0 13.5cm, width=\columnwidth]{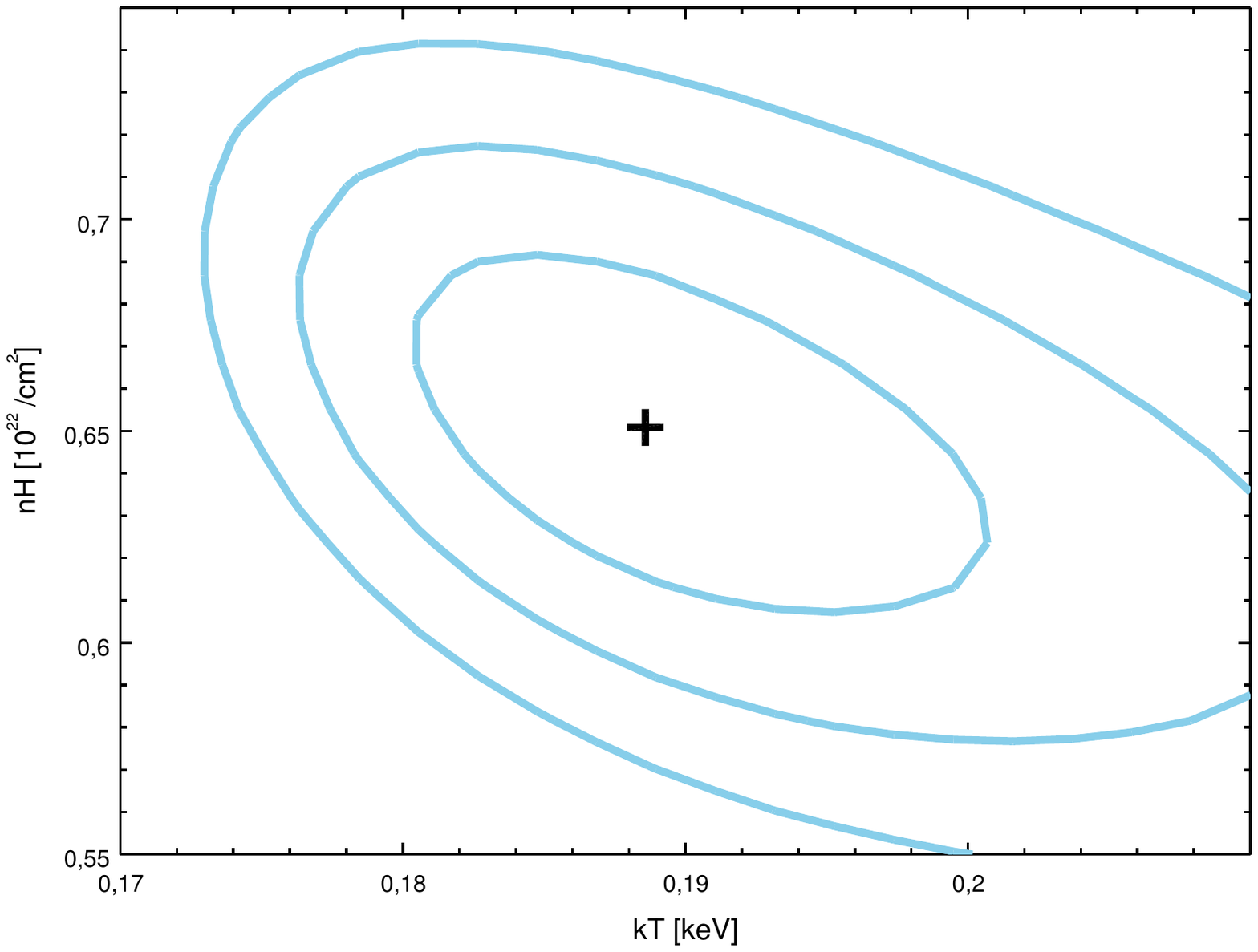}
\includegraphics[trim=0 0 0 13.5cm, width=\columnwidth]{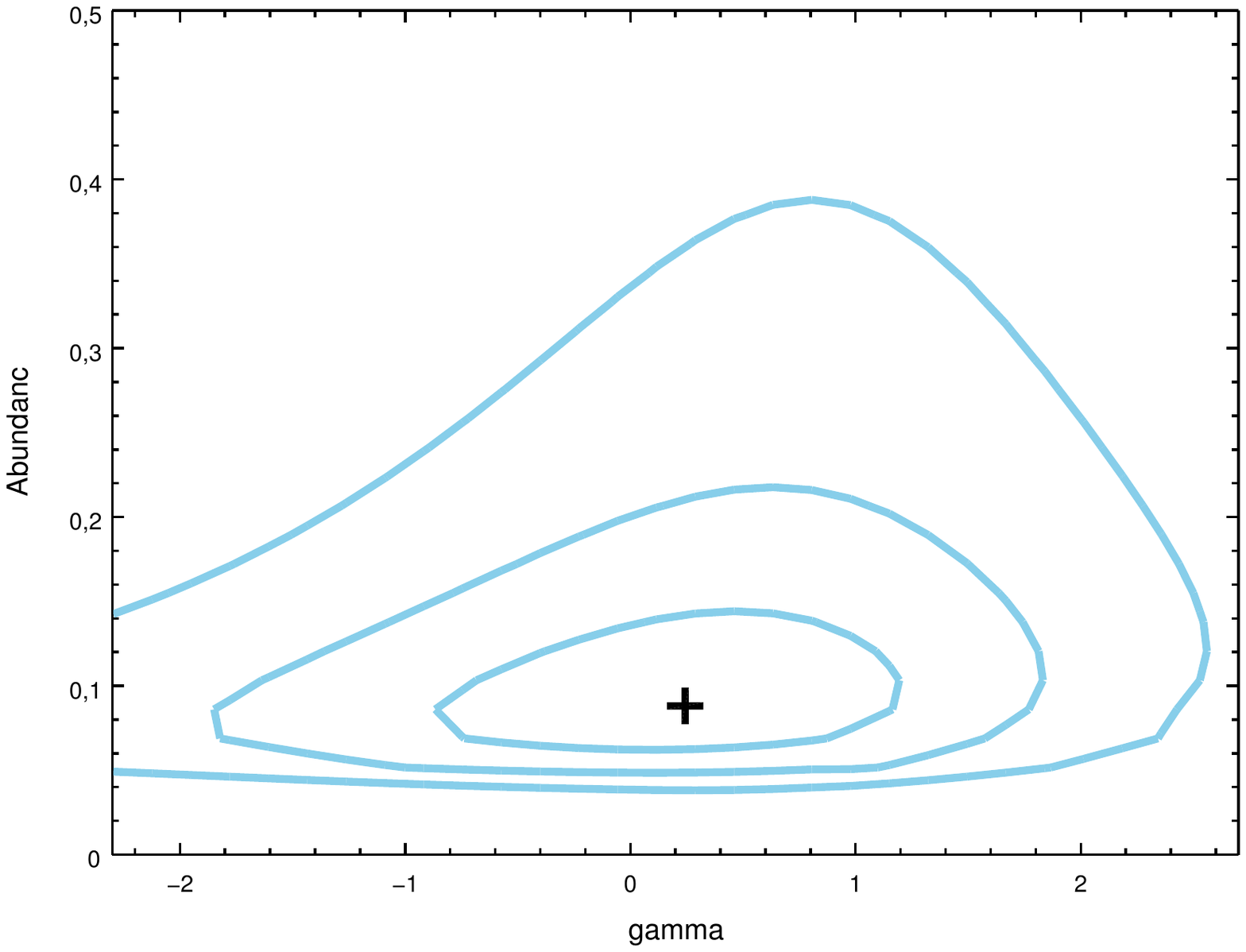}
\includegraphics[trim=0 0 0 13.5cm, width=\columnwidth]{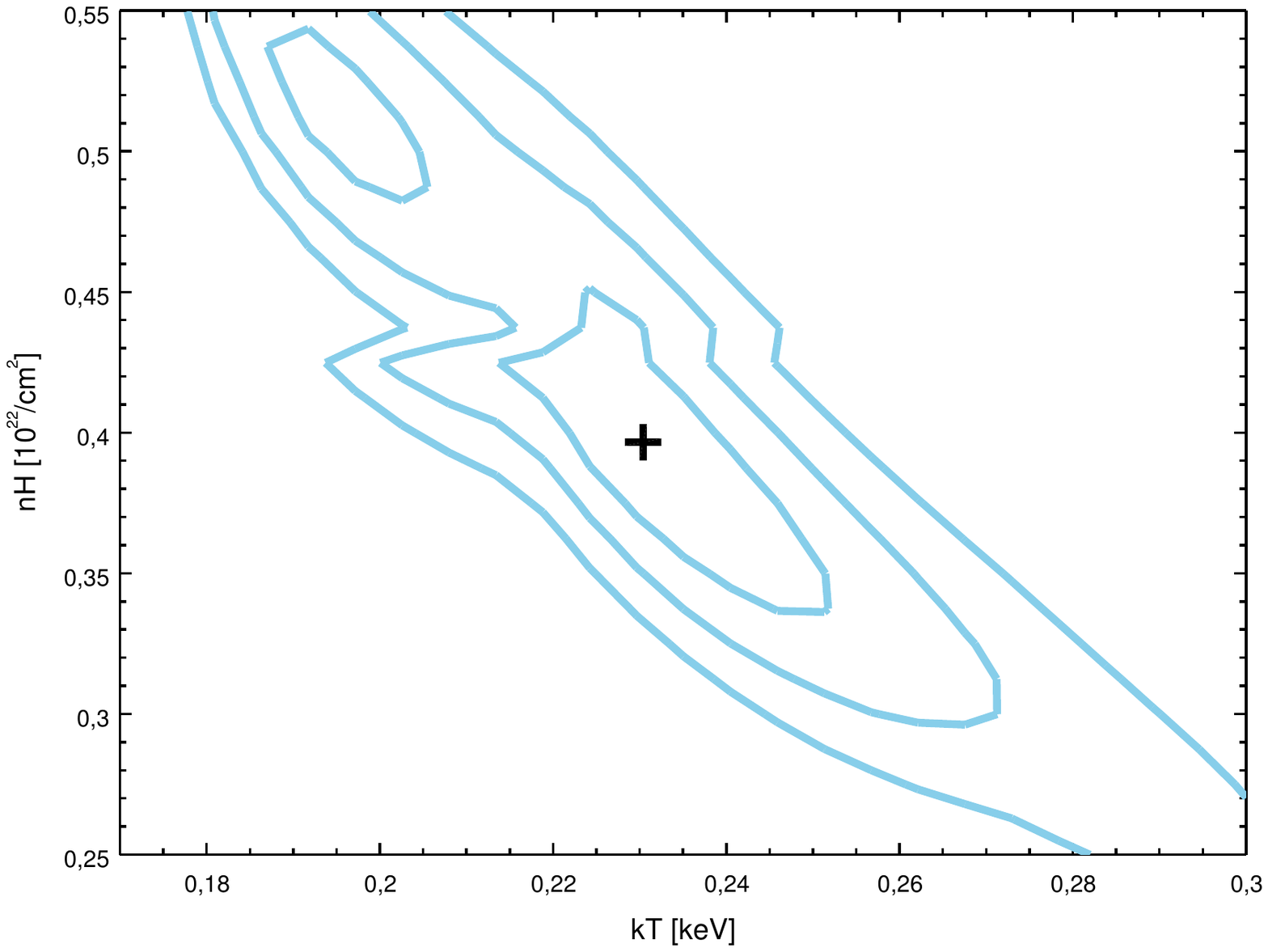}
	\caption{Confidence contours of 1, 2, and 3 $\sigma$ for the {\tt chi2datavar} statistic on the two-thaw-parameters planes, for the West region (upper panels), and the East region (lower panel).}
	\label{fig:contours}
\end{figure*}

The final results obtained with the Levenberg-Marquardt optimalization method, using the chi-squared statistic ({\tt chi2datavar}) with variance calculated from the data, are presented in Figure\,\ref{fig:fit_dw}, and the corresponding best-fit values of the model free parameters are summarized in Table\,\ref{tab:fit}. In Figure\,\ref{fig:contours}, we provide also the confidence contour plots of the main model parameters for the two-component model applied to the East and West regions. The results of the spectral fitting for all four analyzed regions can be summarized as follows:
\begin{itemize}
\item In the East region, which is seemingly displaced from the radio-emitting plasma of the main lobe, we clearly see a thermal component with the best-fit temperature of $kT \sim 0.2$\,keV, and sub-solar abundance $\sim 0.1$, moderated by a hydrogen column density $\sim 0.4\times 10^{22}$\,cm$^{-2}$ much in excess of the Galactic value. The power-law component seen in the spectrum is characterized by a very flat (although not well constrained) slope and a low amplitude.
\item In the West region, which does overlap with the edge of the Southern radio lobe, we see the same thermal component as in the East region ($kT \sim 0.2$\,keV, sub-solar abundance of $\sim 0.2$, hydrogen column density $\sim 0.65\times 10^{22}$\,cm$^{-2}$). However, in addition we also detected a power-law emission characterized by a relatively steep slope ($\Gamma \simeq 1.75 \pm 0.28$) and a high amplitude.
\item In the South region, which is located at larger distances from the core and which corresponds to the Southernmost edge of the radio counter-lobe, we do not see any particularly pronounced thermal emission component, or a hydrogen column density in excess of the Galactic value. The single absorbed power-law component (though with hardly constrained absorbing column density, consistent with zero) dominating the radiative output of the region, is characterized by a relatively steep slope ($\Gamma \simeq 2.14 \pm 0.13$), and a high amplitude.
\item  In the case of the North region, which is also located at a larger distance from the core (when compared to the West and East regions), and which overlaps with the edge of the main radio lobe, despite a very low photon statistics, an acceptable fit could be obtained assuming a single power-law model, yielding a steep slope of the continuum ($\Gamma \simeq 2.02 \pm 0.43$), its relatively small amplitude, and unconstrained absorbing column density consistent with zero. Removal of the internal absorption from the model improves the fit. 
\end{itemize}

As mentioned previously, because of the extremely high flux of the Cen\,A nucleus, leading to a severe pile-up in the detector, we have restricted our spectral analysis to the low-energy range of the ACIS instrument \citep[see in this context][]{Evans04,Kraft07,Croston09}. However, the contribution of the core's PSF seems still to be present within the East region, even below 3\,keV. Indeed, as discussed in \citet{Mingo11} and \citet{Hardcastle16}, the extended wings of high-flux unresolved sources are, in general, expected to manifest as low-amplitude and flat-spectrum power-law components even at larger distances from the targets, due to a combination of the energy-dependent {\it Chandra}'s PSF, and a significant instrumental pile-up affecting predominantly low-energy segments of the targets' spectra. Hence, we believe that the low-amplitude and flat-spectrum power-law component seen within the East region, is due to this effect, i.e. represents only the very broad PSF wings of the Cen\,A nucleus. As such, it should be also present in the West region, located at a comparable distance from the core. We have therefore repeated the spectral fitting for the West region adding an additional \emph{flat} power-law component, with the model parameters fixed at the values within $1\sigma$ errors of the best-fit power-law component emerging from the spectral analysis of the East region. The remaining model parameters obtained in this way turned out basically the same as the ones reported in Table\,\ref{tab:fit}, as expected keeping in mind a much lower amplitude of the flat-spectrum power-law component to be compared with the steep-spectrum one.
 
On the other hand, the steep-spectrum power-law component with the best-fit photon index $\Gamma \sim 2.0$, which can be seen in the North, West, and the South regions, cannot be simply an instrumental artefact. Our fitting results for the South region are in fact in agreement with the modelling presented in \citet{Croston09}, who argued that the X-ray power-law continuum of that region, with the best-fit photon index of $\Gamma \simeq 2.01^{+0.04}_{-0.03}$ (model IV therein), is due to the synchrotron emission of very high-energy electrons energized at the front of the bow shock, induced in the ambient medium by the expanding radio counter-lobe. 

We also note that the field partly overlapping with our West region was also analyzed by \citet[``Region 3'' therein]{Croston09}; the best fit obtained by these authors assumed in this case a thermal model, and yielded a relatively high plasma temperature of $\simeq 0.95$\,keV. The reason for the discrepancy between our fitting results and those presented by \citeauthor{Croston09} for the West feature, is the differences in the source and background extraction regions, as well as in the fitting procedure; as a result, in our spectral modelling instead of a hot plasma we see a prominent power-law component in addition to the relatively cold thermal gas emission. We believe that our estimates are however robust, because almost exactly same thermal gas parameters emerge for the East region, where we do not see any physical power-law component (above the low-level emission related to the PSF wings of the bright nucleus).

The two issues should be clarified at this point regarding the thermal component fitting. One is that the emerging low abundance values may not necessarily correspond to a real low metallicity, but instead may only signal an unresolved multi-temperature gas, for which the best-fit temperature of $\sim 0.2$\,keV should be considered as an average value, while the metallicity is most likely much closer to the Solar one \citep[see in this context the discussion in][regarding the X-ray knots within the northern middle lobe]{Kraft09}. The other issue is that, the residuals of the best-fit models presented in Figure\,\ref{fig:fit_dw} for the East and West regions, indicate clearly the presence of line-like emission features in the analyzed spectra. We have therefore repeated the spectral fitting for both regions adding additional gaussian components; the results of the modelling are given in the Appendix\,\ref{S:Appendix}. In short, by introducing two additional gaussian spectral components, we do improve the quality of the fitting in terms of the reduced $\chi^2$ values. Moreover, the best-fit line positions, $\sim 1$\,keV and $\sim 1.85$\,keV, correspond to the well-known neon/iron L-shell and silicon blends, respectively \citep[see, e.g.,][]{Peterson06,Boehringer10}. We note that previously, \citet{Evans04} reported on the detection of the neutral silicon K$\alpha$ line in the nuclear spectrum of Cen\,A, with the source extracted region corresponding effectively to $\lesssim 1$\,kpc radius \citep[see also the related discussion in][]{Croston09}.

\section{Discussion}
\label{S:discussion}

 One of the two main results emerging from the analysis presented in the previous section, is the detection of the power-law emission component with a relatively steep photon index $\Gamma \sim 2$ not only at the Southernmost edge of the radio counter-lobe, as reported previously by \citet{Croston09}, but also in the North and West regions, which overlap with the side edges of the main radio lobe and the counter-lobe, respectively. We follow \citeauthor{Croston09} in interpreting this emission component as representing the synchrotron continuum of very high-energy electrons energised at the front of the lobes' termination shock. If that is the case indeed, then the implication of our analysis would be that the efficiency of the electron acceleration at the termination shock front, does not vary dramatically over the inner lobes' extension. The acceleration efficiency in this context stands not only for the maximum electron Lorentz factors $\gamma \sim 10^8$ enabled by a balance between the acceleration and radiative cooling timescales \citep[see the discussion in][]{Croston09}, but regards also the potential for the formation of a ``shock-type'' energy spectrum of ultra-relativistic electrons $\propto E^{-s}$ with $s \sim 2$, assuming the observed synchrotron X-ray photons are produced in the strong cooling regime, i.e. that $s = 2(\Gamma -1)$.

The other main finding following from the spectral analysis presented in the previous section, is the presence of a relatively cold and dense gas within the two regions located at kpc distances to the West and East from the nucleus. The exact emission volumes for these are unknown, and in fact hard to define, but based on the particular regions selected for spectral modelling, we estimate such volumes as $V \simeq 0.43$\,kpc$^3$ and $\simeq 0.23$\,kpc$^3$, respectively, assuming the structures are cone-shaped, with the main axes on the plane of the sky. Assuming further that each region is filled uniformly with fully ionized hydrogen, we derive the gas density $n$, total mass $M= m_p n V$, and the internal energy $\varepsilon_{\rm int} = \frac{3}{2} p V$ for the corresponding thermal pressure $p = n kT$, based  on the normalization parameter in the {\tt xsapec} model, $10^{-14} n^2 V/4\pi D^2$\,cgs (see Table\,\ref{tab:fit}), where $D=3.85$\,Mpc is the distance to the source. For the West region, the derived parameters are $n \sim 0.5$\,cm$^{-3}$, $M \sim 5.5 \times 10^6 M_{\odot}$, $p=1.6 \times 10^{-10}$\,dyn\,cm$^{-2}$, and $\varepsilon_{\rm int} \sim 3 \times 10^{54}$\,erg. For the East structures, we obtain $n \sim 0.3$\,cm$^{-3}$, $M \sim 1.6 \times 10^6 M_{\odot}$, $p=1.2 \times 10^{-10}$\,dyn\,cm$^{-2}$, and $\varepsilon_{\rm int} \sim 1 \times 10^{54}$\,erg. 

Note that by assuming a clumpy distribution of the X-ray emitting gas in the analyzed regions, the derived gas density, and hence also the pressure, would increase. Such an increase would however be problematic, since even with the filling factor of the order of unity, the thermal gas present in the analyzed regions appears much denser than, and over-pressured with respect to, the diffuse ISM at the corresponding distance from the nucleus; for this ``unperturbed'' ISM, following \citet{Kraft03} hereafter we adopt $n_{\rm ISM} \sim 0.01$\,cm$^{-3}$, $kT_{\rm ISM} \sim 0.35$\,keV, and $p_{\rm ISM} \sim 10^{-11}$\,dyn\,cm$^{-2}$. At the same time, the thermal gas we see in the East and West regions turns out to be in a pressure balance with the non-thermal plasma present around the edges of the lobes, for which \citet{Croston09} estimated $p_{\rm shell} \sim 10^{-10}$\,dyn\,cm$^{-2}$. This could suggest that what we see is simply a result of a shock compression of the ISM by the expanding radio lobes. However, the corresponding high density contrast $n/n_{\rm ISM} \sim 50$, together with the temperature ratio $T/T_{\rm ISM} \lesssim 1$, impose a general problem for \emph{any} interpretation involving adiabatically shocked ISM, in which case one would expect the density jump $\sim 4$ at most, and $T/T_{\rm ISM} \gg 1$. 

An unusual density increase along with a rapid temperature drop down to the pre-shock value, on the other hand, could possibly be encountered in a radiative shock, i.e. when the gas cooling in the near downstream is sufficiently fast that a relatively narrow radiative relaxation layer is formed. For such, in the specific case of an ``isothermal shock'' with the upstream plasma bulk velocity $u_-$, the gas temperature in the far downstream settles at $T \sim T_{\rm ISM}$, and the density contrast reaches
\begin{equation}
n \sim n_{\rm ISM} \, \left(\frac{u_-}{c_{\rm s}}\right)^2 \sim 0.2 \, \left(\frac{u_-}{10^3\,{\rm km\,s^{-1}}}\right)^2 \, {\rm cm^{-3}} \, ,
\label{eq:jump}
\end{equation}
consistently with the gas temperature and density derived above for the analyzed East and West regions, as long as $u_- \sim 10^3$\,km\,s$^{-1}$.

However, the problem with this scenario is that the radiative cooling of the ISM gas immediately behind the shock (related to the free-free and line emission), is not sufficiently short.  In fact, assuming a strong shock with $u_{-} \gg c_{\rm s}$, the compression ratio in the near downstream should be simply $n_{+} \approx 4 \, n_{\rm ISM} \sim 0.04$\,cm$^{-3}$, and the gas temperature
\begin{equation}
kT_{+} = \frac{3}{16} m_p u_{-}^2 \sim 2 \, \left(\frac{u_-}{10^3\,{\rm km\,s^{-1}}}\right)^2 \, {\rm keV} \, .
\end{equation}
For such, with $u_- \sim 10^3$\,km\,s$^{-1}$ the thermal cooling timescale 
\begin{equation}
\tau_{\rm cool} = \frac{5}{2}\,\frac{n kT_{+} }{n^2\Lambda} \sim 300\,{\rm Myr} \, ,
\label{eq:cool}
\end{equation}
where $\Lambda$ stands for the radiative cooling function \citep{Peterson06}, and we assumed approximately one-third solar abundance. This would be then orders of magnitude longer than the dynamical timescale involved,
\begin{equation}
\tau_{\rm dyn} \simeq \frac{d}{u_-} \sim 3\,{\rm Myr} \, ,
\end{equation}
where $d \simeq 2.7$\,kpc is the physical distance from the outer edges of the East and West regions to the Cen\,A nucleus (assuming no significant projection effects).

The emerging relatively low temperature of the thermal gas derived for the analyzed West and East features, could neither be explained --- still in the framework of the shock scenario --- by postulating that the electron temperature is lower than expected from the standard shock jump conditions, because of an inefficient heating of thermal-pool electrons at the shock front. That is due to two main reasons. First, because the decreased efficiency of the electron heating is expected rather at low-Mach number shocks \citep[see the discussion in][and references therein]{Stawarz14}. And second, even if  the fraction of the kinetic energy of the outflow that is dissipated to heat thermal electrons at the shock front is in our case low, the electron--ion temperature equilibration due to Coulomb collisions would anyway increase the electrons temperatures in the near downstream of the shock, on the particularly short timescale of
\begin{equation}
\tau_{\rm ei} \sim 0.01 \, \frac{m_p \, (kT_e)^{3/2}}{e^4 n \sqrt{m_e}} \lesssim 0.01\,{\rm Myr} \, ,
\end{equation}
assuming electron temperature $kT_e \sim 0.2$\,keV and number density as given in equation\,\ref{eq:jump}.

The alternative explanation for the East and West structures, motivated by their cone-shape morphologies, as well as the derived temperatures which are comparable to the ISM temperature, could be that those features represent Mach cones, formed after the ejections of dense plasmoids from the nucleus of Cen\,A, with supersonic velocities. Such ejections should drive the pressure waves, which merge at the Mach cone into a sonic boom propagating further within the ISM at the speed of sound $c_{\rm s}= \sqrt{5 kT_{\rm ISM}/3 m_p} \sim 230$\,km\,s$^{-1}$. In the framework of this scenario, the velocity of the ejection, $u_{\rm ej}$, could be estimated by measuring on the {\it Chandra} maps the half-opening angles of the East and West cones, $\theta$, which give the Mach numbers $ \mathcal{M} = 1/\sin\theta$. For both analyzed regions we obtain roughly $\mathcal{M} \simeq 3-5$, i.e. the ejection velocity within the range $u_{\rm ej} = \mathcal{M} \,  c_{\rm s} \sim 600-1,000$\,km\,s$^{-1}$. This leads to the elapse time since the ejection event $\tau_{\rm ej} \sim d/ u_{\rm ej}  \sim 3$\,Myr. This timescale is of the same order as the timescale for the formation of the inner radio jets and lobes in Cen\,A \citep[see][and references therein]{Croston09,Morganti10}, and also as the  X-ray cooling timescale of 0.2\,keV gas with 0.3\,cm$^{-3}$ number density (cf. equation\,\ref{eq:cool}). Hence, the ejection episode postulated here could indeed be considered as accompanying/coinciding with the onset of the currently ongoing jet activity in the system.
 
Moreover, the derived ejection velocities are relatively high, but on the other hand consistent with the velocities of nuclear outflows detected in several AGN. In the particular case of Cen\,A, we note in this context that, based on the {\it Suzaku} observations of the Cen\,A nucleus, \citet{Tombesi14} claimed the detection of the Fe\,XXV He$\alpha$ and Fe\,XXVI Ly$\alpha$ absorption lines with the equivalent widths of the order of 10\,eV, corresponding to the ionized hot absorber with the outflow velocities $\leq 1,500$\,km\,s$^{-1}$. What is more, in the recent {\it Herschel} data for the central 500\,pc of Cen\,A, \citet{Israel17} detected an outflow of cold, neutral and ionized gas, roughly along the axis of the radio jet, with a mass of several million solar masses, and the \emph{projected} velocity of 60\,km\,s$^{-1}$. We do not necessarily identify those currently observed nuclear or circum-nuclear outflows with the ejection episode postulated here to explain the East and West features at kpc distances from the core. The point is, rather, that one can indeed expect formation of massive gaseous outflows and plasma ejections in the system, with high and very high velocities. 

A possible complication to the above-drafted scenario could be, however, again the very high pressure contrast between the analyzed regions and the ISM. On the other hand, as the required propagation velocity is most likely super-Alfvenic, for the Alfven speed $v_A = B_{\rm ISM} / \sqrt{4 \pi m_p n_{\rm ISM}} < 200$\,km\,s$^{-1}$ with $n_{\rm ISM} \sim 0.01$\,cm$^{-3}$ and the anticipated $B_{\rm ISM} < 10$\,$\mu$G, one may speculate that the magnetic draping effect, expected as long as the coherence scale of the ISM magnetic field is large enough \citep[see in this context][]{Moss96}, effectively increases the total ISM pressure ahead of the ejection by the piled-up magnetic field \citep[see][]{Lyutikov06}, so that the pressure balance is maintained.

\section{Summary and final remarks}
\label{S:conclusions}

In this paper we re-analyze the archival {\it Chandra} data for the central parts of the Centaurus\,A radio galaxy, aiming for a systematic investigation of the X-ray emission associated with the inner radio lobes, and their immediate surroundings. After inspection of the X-ray hardness maps of the system, we focus on four distinct features characterized by the soft excess with respect to the adjacent fields. Those include the two regions located at kpc distances from the nucleus to the West and East, the extended bow-shock structure to the South, and a fragment of a thin arc North from the center. The selected North, West, and South features coincide with the edges of the radio lobes, while the East structure is seemingly displaced from the radio-emitting plasma. 

We perform the spectral analysis for the selected regions, assuming a combination of the absorbed power-law and thermal emission components. We found out that for the North and South features, a simple power-law model consistent with no thermal contribution and no intrinsic absorption, provided satisfactory fits to the data. The spectra of the East and West regions, on the other hand, could not be fitted at all with a one-component model, consisting of either an absorbed single power-law emission, or an absorbed single-temperature plasma; for those, a two-component model was indeed required.

One of the two main results emerging from our spectral analysis, is the detection of the power-law emission component with a relatively steep photon index $\Gamma \sim 2$ not only at the Southernmost edge of the radio counter-lobe, as reported previously by \citet{Croston09}, but also in the North and West regions, which overlap with the side edges of the main radio lobe and the counter-lobe, respectively. This emission component can be naturally explained as representing the synchrotron continuum of very high-energy electrons energised at the front of the lobes' termination shock. Hence, we conclude that the efficiency of the electron acceleration at the termination shock front, does not vary dramatically over the inner lobes' extension.

The other main finding following from our spectral analysis, is the presence of a relatively cold (temperature $\simeq 0.2$\,keV) and dense (number density $\sim 0.3$\,cm$^{-3}$) gas within the two regions located at kpc distances to the West and East from the nucleus, which appears over-pressured (by one order of magnitude) with respect to the surrounding diffuse ISM. We argue that the scenario in which this gas represents the ISM shocked by the expanding radio lobes, is not self-consistent, because of the required effectively ``isothermal compression''. Instead, we propose that the presence of such a cold and dense gas could possibly be related to a massive nuclear outflow from the central regions of the galaxy.

\acknowledgments

D.K., V.M., and \L .S. were supported by Polish NSC grant 2016/22/E/ST9/00061. The authors thank the anonymous referee for her/his critical comments and suggestions, which helped to improve the paper substantially. The authors are also grateful to Arti Goyal, for useful discussions on the low-frequency radio maps of Cen A. 
\vspace{5mm}
\facilities{Chandra (ACIS)}
\software{CIAO \citep{Fruscione06}, Sherpa \citep{Freeman01}}

\bibliographystyle{aasjournal}

\begin{thebibliography}{99} 

\bibitem[Abdalla et al.(2018)]{Abdalla18} Abdalla, H., Abramowski, A., et al.\ 2018, \aap, 619, A71.

\bibitem[Abdo et al.(2010a)]{Abdo10a} Abdo, A.~A., Ackermann, M., Ajello, M., et al.\ 2010a, Science, 328, 725 

\bibitem[Abdo et al.(2010b)]{Abdo10b} Abdo, A.~A., Ackermann, M., Ajello, M., et al.\ 2010b, \apj, 719, 1433.

\bibitem[Aharonian et al.(2009)]{Aharonian09} Aharonian, F., Akhperjanian, A.~G., Anton, G., et al.\ 2009, \apj, 695, L40.

\bibitem[B{\"o}hringer \& Werner(2010)]{Boehringer10} B{\"o}hringer, H. \& Werner, N.\ 2010, \aapr, 18, 127

\bibitem[Cappellari et al.(2009)]{Cappellari09} Cappellari, M., Neumayer, N., Reunanen, J., et al.\ 2009, \mnras, 394, 660 

\bibitem[Condon et al.(1996)]{Condon96} Condon, J.~J., Helou, G., Sanders, D.~B., et al.\ 1996, \apjs, 103, 81

\bibitem[Crockett et al.(2012)]{Crockett12} Crockett, R.~M., Shabala, S.~S., Kaviraj, S., et al.\ 2012, \mnras, 421, 1603

\bibitem[Croston et al.(2009)]{Croston09} Croston, J.~H., Kraft, R.~P., Hardcastle, M.~J., et al.\ 2009, \mnras, 395, 1999.

\bibitem[Evans et al.(2004)]{Evans04} Evans, D.~A., Kraft, R.~P., Worrall, D.~M., et al.\ 2004, \apj, 612, 786.

\bibitem[Feain et al.(2011)]{Feain11} Feain, I.~J., Cornwell, T.~J., Ekers, R.~D., et al.\ 2011, \apj, 740, 17 

\bibitem[Ferrarese et al.(2007)]{Ferrarese07} Ferrarese, L., Mould, J.~R., Stetson, P.~B., et al.\ 2007, \apj, 654, 186 

\bibitem[Freeman et al.(2001)]{Freeman01} Freeman, P., Doe, S., \& Siemiginowska, A.\ 2001, \procspie, 4477, 76

\bibitem[Fruscione et al.(2006)]{Fruscione06} Fruscione, A., McDowell, J.~C., Allen, G.~E., et al.\ 2006, \procspie, 6270, 62701V

\bibitem[Hardcastle et al.(2007)]{Hardcastle07} Hardcastle, M.~J., Kraft, R.~P., Sivakoff, G.~R., et al.\ 2007, \apj, 670, L81.

\bibitem[Hardcastle et al.(2009)]{Hardcastle09} Hardcastle, M.~J., Cheung, C.~C., Feain, I.~J., \& Stawarz, {\L}.\ 2009, \mnras, 393, 1041 

\bibitem[Hardcastle et al.(2016)]{Hardcastle16} Hardcastle, M.~J., Lenc, E., Birkinshaw, M., et al.\ 2016, \mnras, 455, 3526

\bibitem[Israel et al.(2017)]{Israel17} Israel, F.~P., G{\"u}sten, R., Meijerink, R., et al.\ 2017, \aap, 599, A53

\bibitem[Israel(1998)]{Israel98} Israel, F.~P.\ 1998, \aapr, 8, 237 

\bibitem[Kalberla et al.(2005)]{Kalberla05} Kalberla, P.~M.~W., Burton, W.~B., Hartmann, D., et al.\ 2005, \aap, 440, 775

\bibitem[Karovska et al.(2002)]{Karovska02} Karovska, M., Fabbiano, G., Nicastro, F., et al.\ 2002, \apj, 577, 114.

\bibitem[Kataoka et al.(2006)]{Kataoka06} Kataoka, J., Stawarz, {\L}., Aharonian, F., et al.\ 2006, \apj, 641, 158.

\bibitem[Kraft et al.(2003)]{Kraft03} Kraft, R.~P., V{\'a}zquez, S.~E., Forman, W.~R., et al.\ 2003, \apj, 592, 129

\bibitem[Kraft et al.(2007)]{Kraft07} Kraft, R.~P., Nulsen, P.~E.~J., Birkinshaw, M., et al.\ 2007, \apj, 665, 1129

\bibitem[Kraft et al.(2009)]{Kraft09} Kraft, R.~P., Forman, W.~R., Hardcastle, M.~J., et al.\ 2009, \apj, 698, 2036

\bibitem[Lyutikov(2006)]{Lyutikov06} Lyutikov, M.\ 2006, \mnras, 373, 73

\bibitem[Marconi et al.(2006)]{Marconi06} Marconi, A., Pastorini, G., Pacini, F., et al.\ 2006, \aap, 448, 921 

\bibitem[McKinley et al.(2013)]{McKinley13} McKinley, B., Briggs, F., Gaensler, B.~M., et al.\ 2013, \mnras, 436, 1286 

\bibitem[Mingo et al.(2011)]{Mingo11} Mingo, B., Hardcastle, M.~J., Croston, J.~H., et al.\ 2011, \apj, 731, 21


\bibitem[Morganti et al.(1999)]{Morganti99} Morganti, R., Killeen, N.~E.~B., Ekers, R.~D., et al.\ 1999, \mnras, 307, 750

\bibitem[Morganti(2010)]{Morganti10} Morganti, R.\ 2010, PASA, 27, 463 

\bibitem[Moss \& Shukurov(1996)]{Moss96} Moss, D., \& Shukurov, A.\ 1996, \mnras, 279, 229

\bibitem[Neff et al.(2015)]{Neff15} Neff, S.~G., Eilek, J.~A., \& Owen, F.~N.\ 2015, \apj, 802, 88

\bibitem[Neumayer et al.(2007)]{Neumayer07} Neumayer, N., Cappellari, M., Reunanen, J., et al.\ 2007, \apj, 671, 1329 

\bibitem[Oosterloo \& Morganti(2005)]{Oosterloo05} Oosterloo, T.~A., \& Morganti, R.\ 2005, \aap, 429, 469

\bibitem[Peterson, \& Fabian(2006)]{Peterson06} Peterson, J.~R., \& Fabian, A.~C.\ 2006, \physrep, 427, 1

\bibitem[Rejkuba(2004)]{Rejkuba04} Rejkuba, M.\ 2004, \aap, 413, 903 

\bibitem[Sahakyan et al.(2013)]{Sahakyan13} Sahakyan, N., Yang, R., Aharonian, F.~A., et al.\ 2013, \apj, 770, L6.

\bibitem[Salom{\'e} et al.(2016)]{Salome16} Salom{\'e}, Q., Salom{\'e}, P., Combes, F., et al.\ 2016, \aap, 586, A45

\bibitem[Sanchez et al.(2018)]{Sanchez18} Sanchez, D., Holler, M., Taylor, A. et al. for the HESS Collaboration 2018, talk during TeVPA2018 (Berlin)  
	
\bibitem[Snios et al.(2019)]{Snios19} Snios, B., Wykes, S., Nulsen, P.~E.~J., et al.\ 2019, \apj, 871, 248.

\bibitem[Stawarz et al.(2014)]{Stawarz14} Stawarz, {\L}., Szostek, A., Cheung, C.~C., et al.\ 2014, \apj, 794, 164

\bibitem[Stawarz et al.(2013)]{Stawarz13} Stawarz, {\L}., Tanaka, Y.~T., Madejski, G., et al.\ 2013, \apj, 766, 48 

\bibitem[Steinle et al.(1998)]{Steinle98} Steinle, H., Bennett, K., Bloemen, H., et al.\ 1998, \aap, 330, 97.

\bibitem[Sun et al.(2016)]{Sun16} Sun, X.-n., Yang, R.-z., Mckinley, B., \& Aharonian, F.\ 2016, \aap, 595, A29 

\bibitem[Tanada et al.(2019)]{Tanada19} Tanada, K., Kataoka, J., \& Inoue, Y.\ 2019, \apj, 878, 139

\bibitem[Tombesi et al.(2014)]{Tombesi14} Tombesi, F., Tazaki, F., Mushotzky, R.~F., et al.\ 2014, \mnras, 443, 2154

\bibitem[Quillen et al.(2006)]{2006ApJ...645.1092Q} Quillen, A.~C., Brookes, M.~H., Keene, J., et al.\ 2006, \apj, 645, 1092 

\bibitem[Werner et al.(2006)]{Werner06} Werner, N., de Plaa, J., Kaastra, J.~S., et al.\ 2006, \aap, 449, 475


\end{thebibliography}

\appendix
\section{Emission lines in the West and the East regions}
\label{S:Appendix}

\begin{figure*}[!h]
	\centering
	\includegraphics[width=0.45\textwidth]{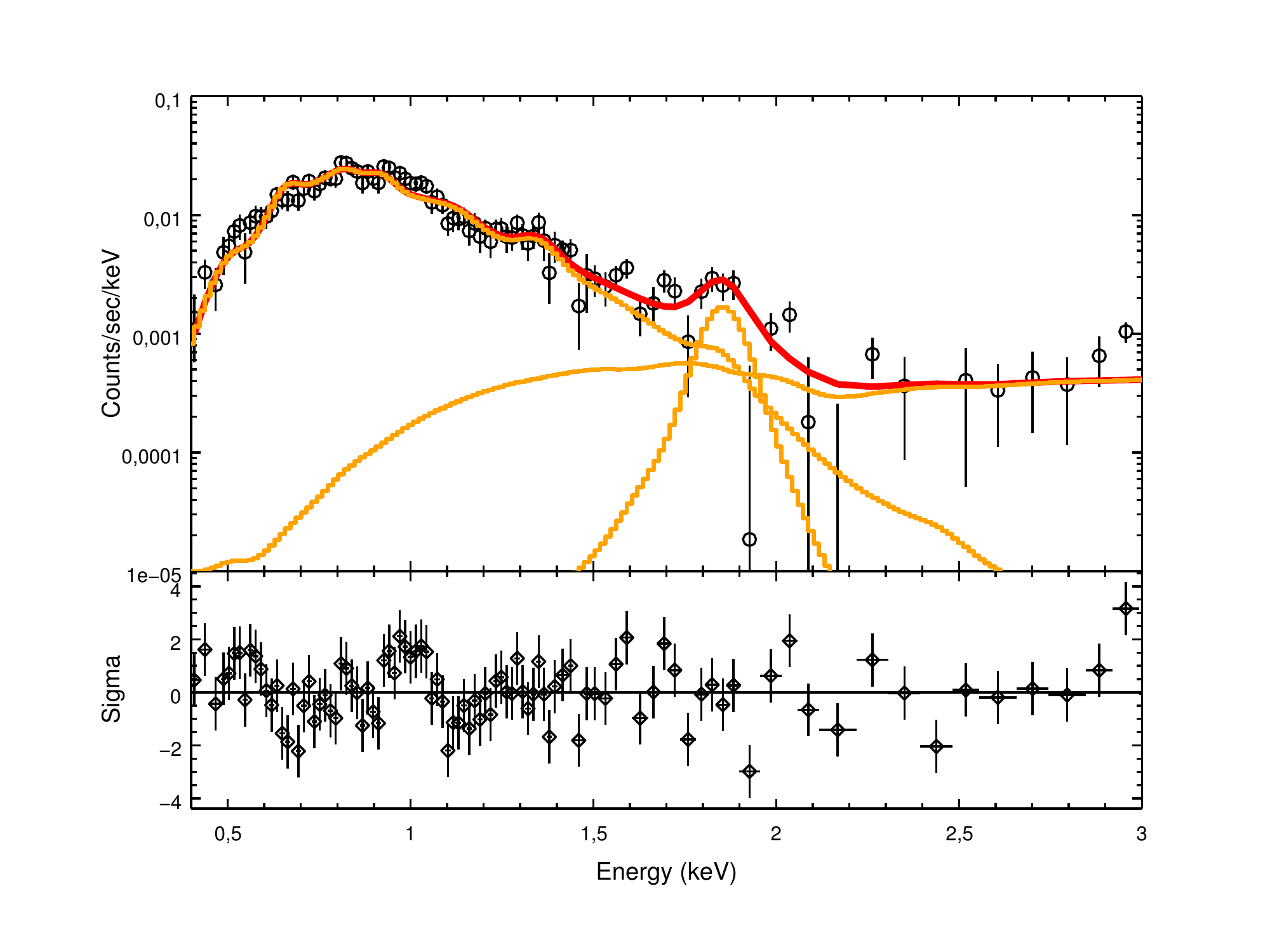}
	\includegraphics[width=0.45\textwidth]{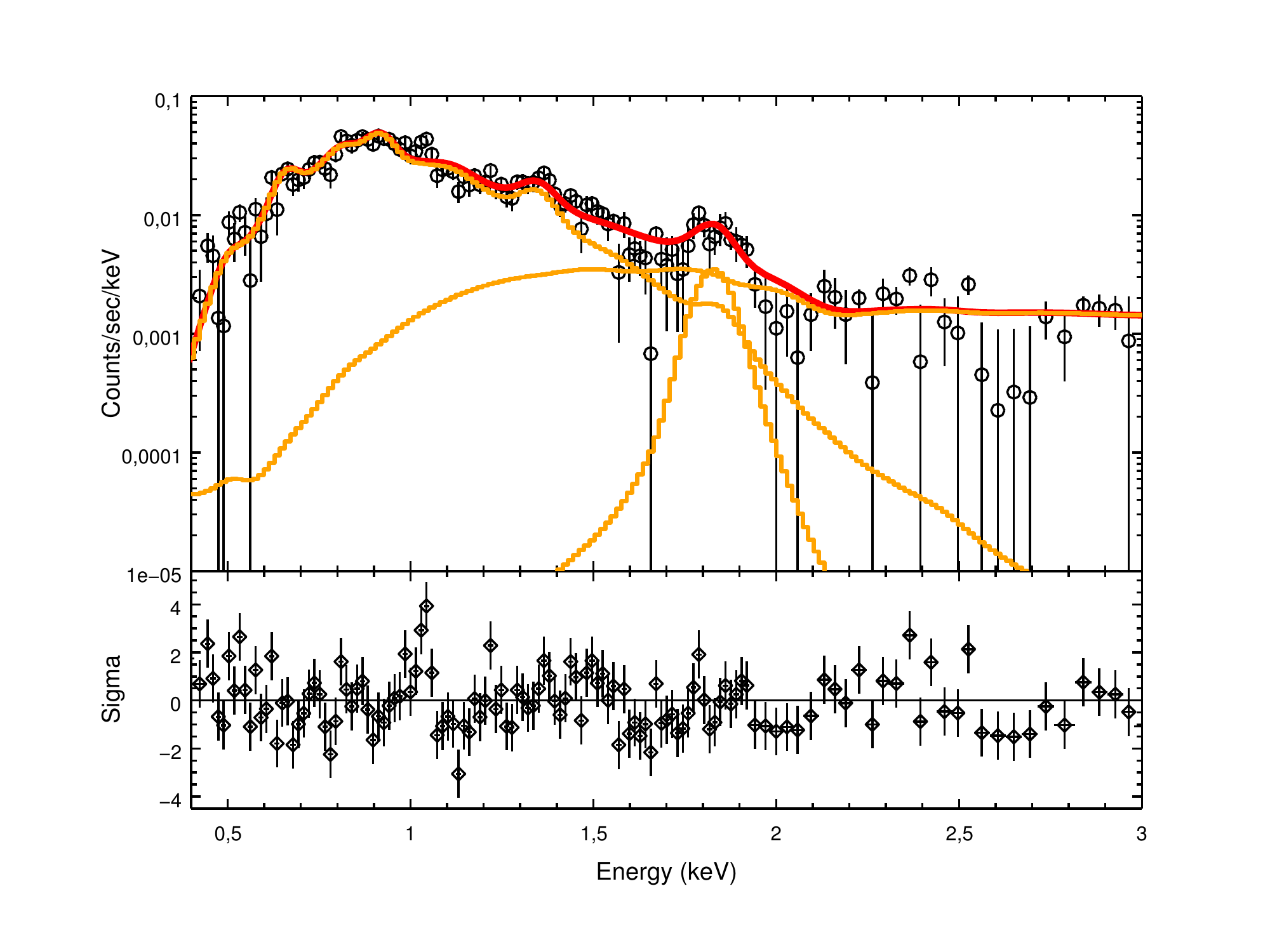}
	\includegraphics[width=0.45\textwidth]{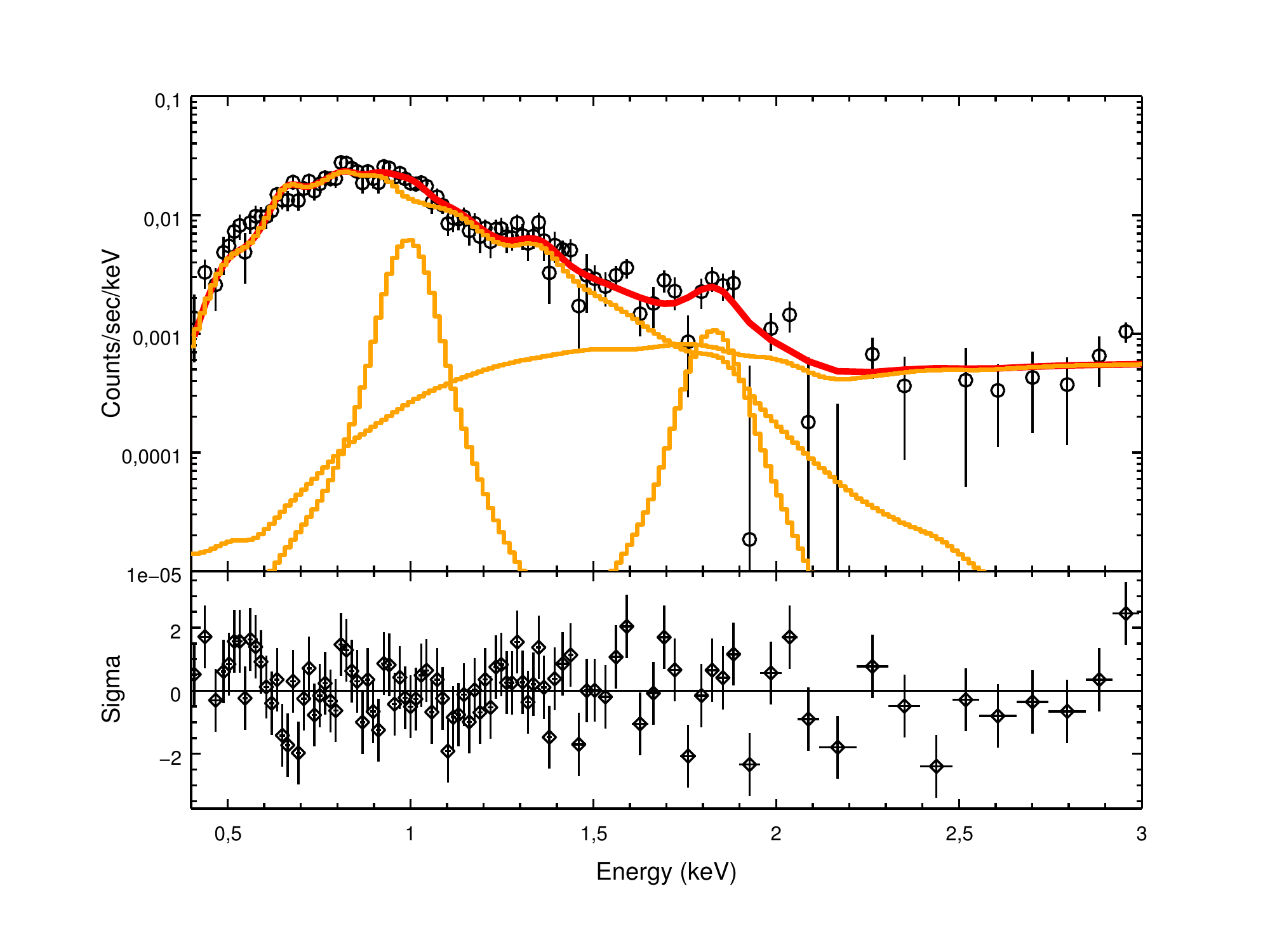}
	\includegraphics[width=0.45\textwidth]{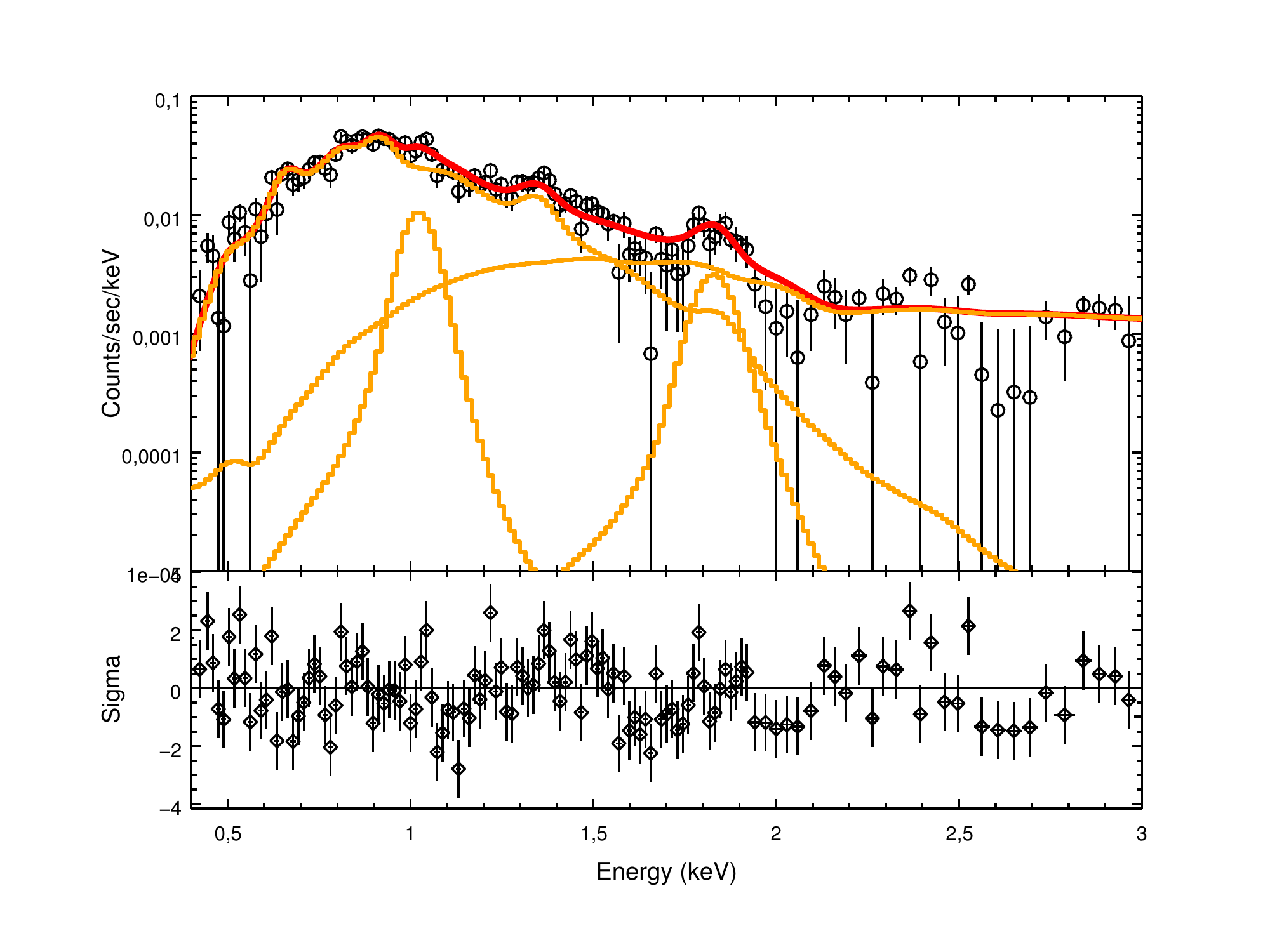}
	\caption{  {\it Chandra} spectra along with the best-fit models (and residuals) for the East region (left column) and West region (right column). The models displayed consist of a mixture of absorbed thermal and non-thermal components, {\tt xsphabs*(xsapec+xspowerlaw)}, with the addition of either one or two gaussian features {\tt xsgaussian} (upper and lower rows, respectively).}
	\label{fig:lines}
\end{figure*}

Residuals are clearly seen in the best-fit models {\tt xsphabs*(xsapec+xspowerlaw)} presented in Figure\,\ref{fig:fit_dw} for the East and West regions, indicating the presence of line-like emission features. We have investigated this issue by introducing first an additional gaussian component to the model, {\tt xsphabs*(xsapec+xspowerlaw + xsgaussian)}, and next allowing for two different gaussian features, {\tt xsphabs*(xsapec+xspowerlaw + xsgaussian1+ xsgaussian2)}, each time with the source frame line widths frozen at $\sigma =10$\,eV, and plasma abundances frozen at the best-fit values emerging from the basic {\tt xsphabs*(xsapec+xspowerlaw)} fits for the two regions (see Table\,\ref{tab:fit}). The results of the modelling are presented in Figure\,\ref{fig:lines}, and summarized in Table\,\ref{tab:fit-G}. As shown, by introducing two additional gaussian spectral components, one can indeed improve the quality of the fitting in terms of the reduced $\chi^2$ values, with no significant change in the best-fit values of the other model parameters. The best-fit positions of the first line-like feature reads as $\sim 1$\,keV, indicating either a hydrogen-like Ne Ly$\alpha$ line, or the iron L-shell blend \citep[the position of which depends however on the plasma temperature; see][]{Boehringer10}; the best-fit position of the other line, $\sim 1.85$\,keV, allows for the identification with the Si XIII blend \citep[e.g.,][]{Peterson06}.

	\begin{table*}[!t]
		{\scriptsize
			\begin{center}
				\caption{Spectral fitting results with additional gaussian components for the East and West regions.}
				\label{tab:fit-G}
				\begin{tabular}{cllll}
					\hline\hline
					Region/Model &  Parameter & Value &  $1\sigma$ errors & Units \\ 
					\hline\hline\\
					{\bf East}  & kT  &  0.22 & 0.01& keV \\ 
					{\tt xsphabs*(xsapec+xspowerlaw} & norm  & $0.5$ & $0.1$ & $10^{-2}\times $\,{\tt apec} \\ 
					{\tt + xsgaussian)} &  $\Gamma$  & 0.11 & 0.50 & --- \\ 
					& ampl  & $1.5$ & $0.6$ & $10^{-6} \times$\,ph/keV/cm$^2$ at 1keV \\  
					& Line position  & 1.85 & 0.01 & keV \\
					&  Line normalization  & 0.93 & 0.02 &$10^{-6} \times$\,ph/cm$^2$/s  \\
					& $N_{\rm H}$  & 0.42 & 0.04& $10^{22}$\,cm$^{-2}$\\ 
					& Final fit statistic & 124.08 & &   \\
					& Degrees of freedom  & 91 & & \\
					\\
					\hline \\

		{\bf West}	& kT  &  0.19 & 0.01& keV \\ 
	{\tt xsphabs*(xsapec+xspowerlaw} & norm  & $2.2$ & $0.5$ & $10^{-2}\times $\,{\tt apec} \\ 
	{\tt + xsgaussian)} & $\Gamma$  & 1.4 & 0.34 & --- \\ 
	& ampl  & $22.1$ & $6.5$ & $10^{-6} \times$\,ph/keV/cm$^2$ at 1keV \\  
	& Line position  & 1.83 & 0.02 & keV \\
	&  Line normalization  & 1.4 & 0.3 &$10^{-6} \times$\,ph/cm$^2$/s  \\
	& $N_{\rm H}$  & 0.67 & 0.02& $10^{22}$\,cm$^{-2}$\\ 
	& Final fit statistic & 190.5 & &   \\
	& Degrees of freedom  & 127 & & \\
	\\
	\hline \hline	\\
					{\bf East} & kT  &  0.22 & 0.01& keV \\ 
					{\tt xsphabs*(xsapec+xspowerlaw} & norm  & $0.4$ & $0.2$ & $10^{-2}\times $\,{\tt apec} \\ 
					{\tt + xsgaussian1+ xsgaussian2)} & $\Gamma$  & 0.25 & 0.74 & --- \\ 
					& ampl  & $2.4$ & $1.65$ & $10^{-6} \times$\,ph/keV/cm$^2$ at 1keV \\  
					& Line 1 position  & 1.0 & --- & keV \\
					&  Line 1 normalization  & 7.6 & 1.6 &$10^{-6} \times$\,ph/cm$^2$/s  \\
					& Line 2 position  & 1.83 & 0.03 & keV \\
					& Line 2 normalization  & 0.42 & 0.21 &$10^{-6} \times$\,ph/cm$^2$/s  \\
					& $N_{\rm H}$  & 0.42 & 0.03& $10^{22}$\,cm$^{-2}$\\ 
					& Final fit statistic & 99.55 & &   \\
					& Degrees of freedom  & 89 & & \\
					\\
					\hline \\
					
					{\bf West}	& kT  &  0.19 & 0.01& keV \\ 
	{\tt xsphabs*(xsapec+xspowerlaw} & norm  & $0.4$ & $0.2$ & $10^{-2}\times $\,{\tt apec} \\ 
	{\tt + xsgaussian1+ xsgaussian2)} & $\Gamma$  & 1.75 & 0.32 & --- \\ 
	& ampl  & $30.2$ & $7.7$ & $10^{-6} \times$\,ph/keV/cm$^2$ at 1keV \\  
	& Line 1 position  & 1.02 & --- & keV \\
	&  Line 1 normalization  & 17.1 & 3.4 &$10^{-6} \times$\,ph/cm$^2$/s  \\
	& Line 2 position  & 1.83 & 0.02 & keV \\
	& Line 2 normalization  & 1.2 & 0.3 &$10^{-6} \times$\,ph/cm$^2$/s  \\
	& $N_{\rm H}$  & 0.42 & 0.03& $10^{22}$\,cm$^{-2}$\\ 
	& Final fit statistic & 99.55 & &   \\
	& Degrees of freedom  & 89 & & \\					
	\\
	\hline\hline

				\end{tabular}
			\end{center}
		}
	\end{table*}

\end{document}